# Artificial Intelligence Enhanced Digital Nucleic Acid Amplification Testing for Precision Medicine and Molecular Diagnostics

Yuanyuan Wei[1, 2], Xianxian Liu[3, 4], Changran Xu[5], Guoxun Zhang[6], Wu Yuan[1], Ho-Pui Ho[1, *], Mingkun Xu[3, 7, *]

[1] Department of Biomedical Engineering, The Chinese University of Hong Kong, Shatin, Hong Kong SAR, 999077, China. E-mail: aaron.ho@cuhk.edu.hk.

[2] Department of Neurology, David Geffen School of Medicine, University of California, Los Angeles, California, 90095, USA.

[3] Guangdong Institute of Intelligence Science and Technology, Hengqin, Zhuhai, 519031, China. E-mail: xumingkun@gdiist.cn.

[4] Department of Computer and Information Science, University of Macau, Macau SAR, 999078, China.

[5] Department of Computer Science Engineering, The Chinese University of Hong Kong, Shatin, Hong Kong SAR, 999077, China.

[6] Department of Automation, Tsinghua University, Beijing, 100084, China.

[7] Center for Brain-Inspired Computing Research (CBICR), Department of Precision Instrument, Tsinghua University, Beijing, 100084, China.

**Abstract**

The precise quantification of nucleic acids is pivotal in molecular biology, underscored by the rising prominence of nucleic acid amplification tests (NAAT) in diagnosing infectious diseases and conducting genomic studies. This review examines recent advancements in digital Polymerase Chain Reaction (dPCR) and digital Loop-mediated Isothermal Amplification (dLAMP), which surpass the limitations of traditional NAAT by offering absolute quantification and enhanced sensitivity. In this review, we summarize the compelling advancements of dNNAT in addressing pressing public health issues, especially during the COVID-19 pandemic. Further, we explore the transformative role of artificial intelligence (AI) in enhancing dNAAT image analysis, which not only improves efficiency and accuracy but also addresses traditional constraints related to cost, complexity, and data interpretation. In encompassing the state-of-the-art (SOTA) development and potential of both software and hardware, the all-encompassing Point-of-Care Testing (POCT) systems cast new light on benefits including higher throughput, label-free detection, and expanded multiplex analyses. While acknowledging the enhancement of AI-enhanced dNAAT technology, this review aims to both fill critical gaps in the existing technologies through comparative assessments and offer a balanced perspective on the current trajectory, including attendant challenges and future directions. Leveraging AI, next-generation dPCR and dLAMP technologies promises integration into clinical practice, improving personalized medicine, real-time epidemic surveillance, and global diagnostic accessibility. Our analyses anticipate enhanced precision and accessibility in point-of-care testing, ushering a shift toward data-driven, adaptive healthcare. These advancements are poised to revolutionize public health policy and invigorate biomedical research, addressing emerging pathogens and complex genetic disorders.

## 1. Introduction

Since the inception of life approximately 3.5 billion years ago, nucleic acids have stood as the fundamental arbiters of genetic information carriers of life's blueprint. Their intricate structure and functions have galvanized scientific endeavors, propelling landmark discoveries that elucidate life's intricacies and origins. In the present day, the increasing prevalence of infectious diseases worldwide, coupled with the unwavering threat of the COVID-19 pandemic which has impacted over 711 million people (704,753,890 confirmed cases and 7,010,681 deaths) globally as of April 2024[1], has accentuated the pressing need for rapid and precise quantification of nucleic acids[2–4]. Considering that each mammalian cell contains only about 6 picograms of DNA, the implementation of ultrasensitive detection methods has never been more paramount. Rapid and accurate molecular diagnostics are crucial in preventing the global pandemic of infectious diseases[5], including malaria[6], dengue[7], tuberculosis[8], and monkeypox[9]. Nucleic Acid Amplification Tests (NAAT) are supremely specific and sensitive, capable of detecting as few as ten copies of the target per reaction[10]. Their status in confirming infectious diseases is cemented by amplifying and pinpointing pathogens' nucleic acids[11,12]. Recent innovations such as digital Polymerase Chain Reaction (dPCR)[13] and digital Loop-mediated Isothermal Amplification (dLAMP)[14,15] have revolutionized the field of nucleic acid quantification. They've led to a shift towards absolute quantification, a technique capable of detecting precise gene expression levels down to a single molecule and enhancing viral particle detection, which is essential in the realms of virology, oncology, and genetic disorders among others[12,16].

dPCR is a technology designed to perform polymerase chain reactions on a massive scale, all within microscopic partitions or "digital" reactions[17]. Unlike traditional PCR which amplifies DNA in a single reaction chamber[18], dPCR divides the DNA sample into thousands or even millions of individual and parallel PCR reactions in monodispersed picoliter microreactors[19,20]. Some of these may contain one or more target DNA molecules, while others may contain no target molecules at all. After amplification, each partition is examined for the presence (positive) or absence (negative) of the amplification product (detected via fluorescence)[21]. The fraction of positive reactions is then used with Poisson statistics to determine the absolute number of target molecules in the original sample[22]. Contrasting with conventional PCR and quantitative PCR (qPCR), dPCR's absolute quantification subverts amplification bias, propelling it beyond the standard curvilinear limitations and boosting reproducibility[23]. Up to 1,000,000 reactors with densities of up to 440,000 reactors $cm^{-2}$ and a dynamic range of $10^7$, dPCR is capable of detecting below one copy per 100,000 wild-type sequences and the discrimination of a 1% difference in chromosome copy number[24].

Accurate identification of positive microreactors in dPCR products thus ensures the reliability of nucleic acid quantification. However, traditional detection methods such as commercialized machines, flow cytometry, and manual fluorescence imaging, reveal significant limitations[25]. Commercialized dPCR platforms such as those from Bio-Rad and Raindance, while precise, pose issues with cost, complexity, limited multiplexing, and time demands. High initial and recurring expenses may deter use, especially in budget-strapped labs[26,27]. While flow cytometry stands as a high-throughput analytical technique for cell analysis and sorting[28], it is less practical for the needs of droplet digital PCR (ddPCR) due to the technical, operational, and interpretational challenges[29]. The nature of ddPCR droplets, which are often oil-based, can pose significant incompatibility for flow cytometers optimized for aqueous samples. Meanwhile, maintaining ddPCR droplet integrity during flow cytometric analysis is challenging due to shear forces that can cause coalescence, breakage, or deformation, resulting in sample loss and inaccurate results. Moreover, flow cytometry may not match the sensitivity and specificity required for detecting low-abundance targets in ddPCR, which can be weak and may overlap with background noise, leading to inaccurate quantification and false positives or negatives. Manual fluorescence image analysis dependent on software or manual counting, although useful in various contexts, presents significant drawbacks in accuracy, efficiency, reproducibility, and scalability[30]. Automated counting and dynamic thresholding struggle with calibration and sensitivity to noise and artifacts, particularly with diverse samples or weak signals, leading to quantification errors that affect the precision of nucleic acid detection in key dPCR processes[29].

Recent advancements in droplet microfluidic analysis have been markedly propelled by the integration of artificial intelligence (AI), evidencing a significant leap in the field's data analysis capabilities[31,32]. Neural networks, with their end-to-end learning and adaptive capacities[33,34], align exquisitely with dPCR[35] and dLAMP[36] image analysis demands. By eliminating explicit feature engineering and embracing an automated learning paradigm, they streamline the process, proving adept at managing complex datasets and optimizing analytical processes, tackling cost, methodological intricacies, processing duration, and error susceptibility. One illustrative study employed convolutional neural networks (CNNs), achieving a remarkable 99.71% accuracy in the classification of positive droplets[37]. The detection was also optimized for few-shot training for higher effectiveness requiring less dataset. By occupying models such as Yolov3, an accuracy of 98.98% was obtained while reducing the labeling time by 70%[38]. Moreover, the strong generalized capability of neuron networks has been validated in various dPCR scenarios. With an accuracy of 96.23% and a swift detection speed of 2.5 seconds, the Deep-qGFP model has been successfully demonstrated on droplet-based, microwell-based, and agarose-based dPCR image datasets[39]. Furthermore, data-specific detections have been effectively executed. These include label-free detection, which reports an overall accuracy of 94.3% using StratoLAMP for human HPV[40], and multiplex detection via a single fluorescent channel, achieving an accuracy of greater than 98%[41]. Such integration of deep learning algorithms not only accelerates the analysis throughput but also fortifies the reliability of dPCR, heralding a transformative era for this technology in clinical diagnostics and molecular biology research.

In light of the aforementioned advancements and associated challenges within digital nucleic acid amplification testing (dNAAT), our study focuses on the integration of dNAAT and AI, exploring how this intersection has propelled biomedical research forward. This systematic review presents a comprehensive examination of state-of-the-art (SOTA) AI-based dPCR and dLAMP detection models, as well as the integration of all-encompassing Point-of-Care Testing (POCT) systems. Our rigorous analysis highlights the SOTA models attain fully automated operations and enhanced accuracy in decoding dNAAT end-point image information. The concurrent improvements in hardware and software have heralded a high-throughput, cost-effective yet robust era in nucleic acid analysis. By scrutinizing existing methods and recognizing the potential of AI models, this review highlights the advancements and potential of AI-enhanced dNAAT technology while identifying critical gaps, challenges, and future directions. As such, our research underscores the pivotal role of this fusion in shaping the future of molecular diagnostics.

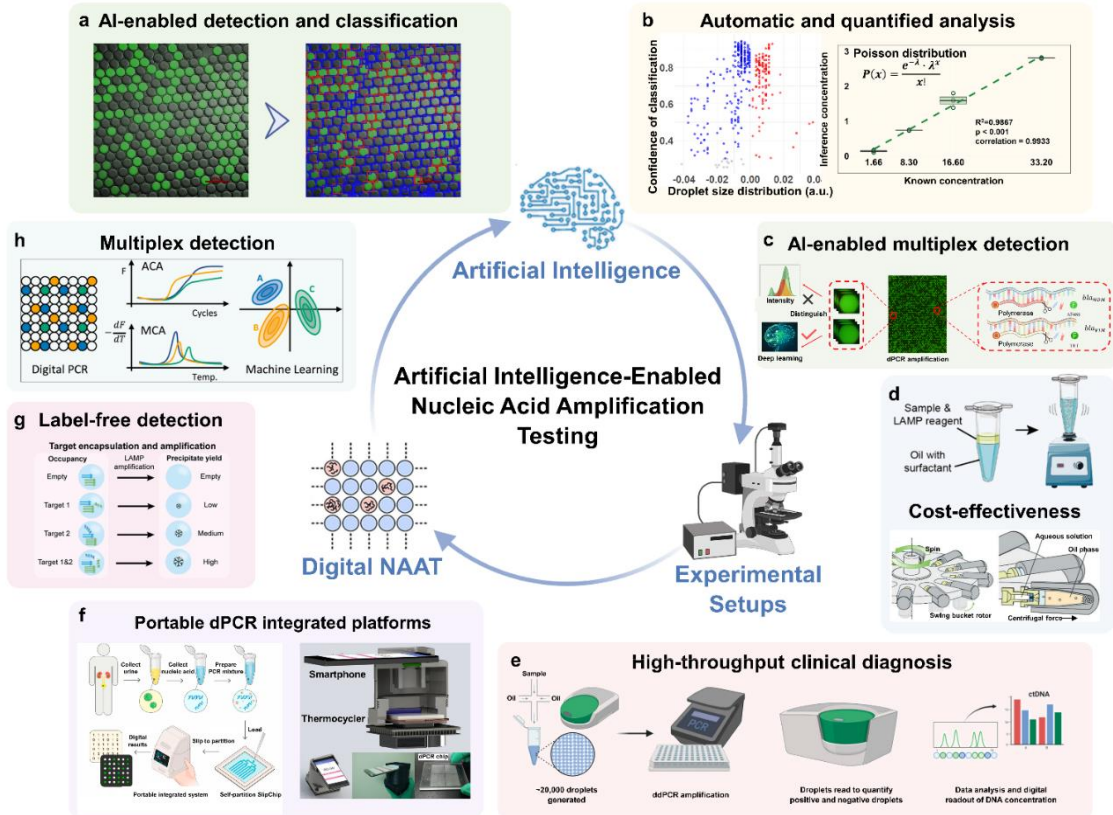

**Fig. 1. Enhancing dNAAT through AI integration and POCT. a.** AI algorithms precisely classify positive (red) and negative (blue) droplets, achieving 96.23% accuracy and quantifying over 2000 microreactors across ten images within 2.5 seconds. Adapted from ref.[39], CC BY 4.0. **b.** Automated dPCR image analysis by AI accurately classifies droplets and infers concentrations using Poisson distribution for enhanced analysis. Adapted from ref.[39], CC BY 4.0. **c.** Deep learning-applied single-color analysis enables multiplex pathogen detection using one fluorescent channel, boosting dNAAT precision and specificity[41]. **d.** Nucleic acid quantification simplified by incorporating common laboratory tools like vortex mixers and centrifuges into the dNAAT procedure. Adapted with permission from the ref[36,42]. CC BY 4.0. and © 2017 Royal Society of Chemistry. **e.** Workflow of high-throughput dPCR in cancer liquid biopsies for clinical detection, based on limiting dilutions, PCR, and Poisson distribution. ddPCR is currently the most commonly used dPCR approach for ctDNA analysis. Adapted with permission from the ref[43]. © 2022 MDPI. **f.** Portability of dPCR demonstrated through handheld, smartphone-compatible devices, reflecting the potential of economical dNAAT for point-of-care use[44]. Innovations include an automated microfluidic dPCR device for self-primed partitioning[45]. **g.** Multiplexed, label-free dLAMP visual stratification by AI eliminates the need for labels, streamlining diagnostics[40]. **h.** Real-time, multiplex dPCR analysis uses trained supervised machine learning models to derive kinetic and thermodynamic insights for robust diagnostics[46]. Each panel illustrates AI's application in dNAAT, from practical techniques to theoretical advancements, portraying a cohesive narrative of technology-enhancing life science methodologies.

## 2. Digital NAAT technology

**Principles and advantages of dNAAT**

dNAAT marks a significant advancement in molecular biology by leveraging compartmentalization to propel nucleic acid quantification into the digital era[47,48]. Unlike traditional methods that provide relative quantification, dNAAT enables the absolute quantification of nucleic acids by partitioning samples into numerous distinct units, thus allowing for the direct counting of target nucleic acid sequences. In dNAAT, samples are divided into thousands of compartments for precise amplification[49–51] as depicted in **Fig. 2a**. These compartments are formed through a variety of partitioning methods, including droplet-based[29,42], chip/chamber-based partitioning[24,52], and agarose-based partitioning[53,54]. Each method follows a systematic process of sample dilution, partitioning, amplification, and detailed analysis. Notably, ddPCR generates picolitre-sized droplets with unmatched precision, demonstrating a coefficient of variation (C.V.) typically below 5%[55]. Meanwhile, microwell dPCR fulfills high-throughput requirements[56,57], and agarose digital PCR offers the benefit of minimizing contamination risks and maintaining sample stability[58].

At the heart of dNAAT lies the innovation of microscale reactors, enabling the digital "digitizing" of target molecules within individual, micron-sized compartments[23]. This structure inherently improves precision in single-cell or molecule analysis, high-throughput screening, quantitative studies due to uniform droplet dispersion, and precise environmental control within each droplet. This not only fosters a fertile ground for robust quantitative studies but also catapults dNAAT into applications spanning intricate environmental analyses and intricate single-molecule studies. Recent advancements integrating these microreactors with real-time detection technologies, such as fluorescence and mass spectrometry, have significantly broadened the potential applications of dNAAT, covering a spectrum from fundamental research to intricate clinical diagnostics.

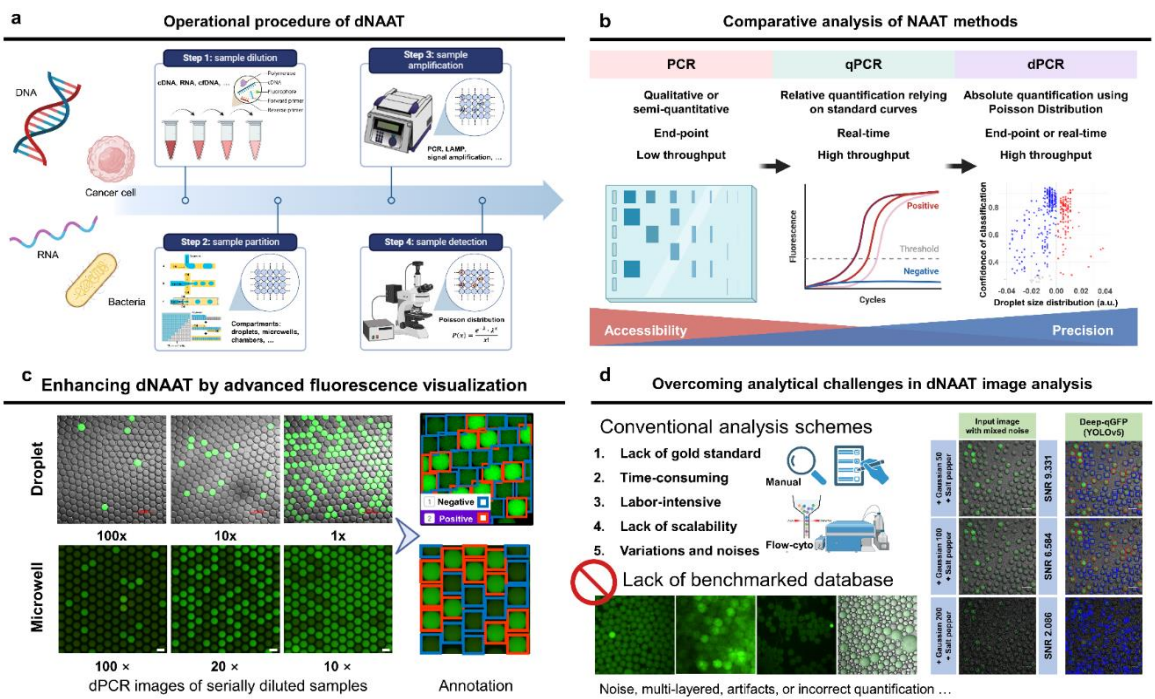

**Fig. 2 Comparative workflow and analytical precision of NAAT techniques. a.** dNAAT operational framework showcases its adaptability in processing diverse sample types (DNA, RNA, cancer cells, bacteria). The workflow includes four phases: (1) Sample Dilution for readying specimens for separation; (2) Sample Partition using droplets, microwells, or agarose for individual isolation; (3) Sample Amplification via PCR, LAMP, or signal enhancement to multiply nucleic acids; and (4) Sample Detection for identifying and quantifying the amplified genetic material. **b.** Comparative analysis of NAAT methods: PCR provides qualitative or semi-quantitative end-point analysis with limited throughput. qPCR offers real-time detection for relative quantification with higher throughput using standard curves. dPCR excels in absolute quantification, enabling both real-time and end-point analyses with high-throughput capabilities and intuitive result interpretation. The transition from PCR to dPCR highlights a shift from general accessibility to enhanced precision, albeit with increased equipment specificity. **c.** Visualization of dNAAT fluorescence in benchtop droplet and microwell dPCR experiments with serially diluted templates, annotated masks, and segmentation into positive (red) and negative (blue) classes by fluorescence intensity. Adapted from ref.[39], CC BY 4.0. **d.** Challenges in dNAAT image analysis: Conventional methods face issues like lack of scalability, need for expertise, subjective interpretation, and extended timelines. AI-driven analysis is challenged by a lack of benchmarked databases and non-ideal imaging conditions, potentially causing inaccuracies from noise, artifacts, or mis-quantification. The impact is demonstrated by applying the YOLOv5 AI model to dPCR images under varied noise levels, highlighting the need for robust AI solutions in molecular diagnostics.

The evolution of PCR techniques, particularly the shift from traditional PCR to qPCR, has been pivotal within the realms of molecular biology, enabling major discoveries across biomedical research domains[59,60]. While the advent of qPCR provided a pathway to real-time detection and relative quantification dependent on standard curves and enhanced throughput, dPCR introduced a revolutionary leap with its capability for absolute quantification, devoid of the necessity for external calibration. It allows for either real-time[25] or end-point analysis, each method providing intuitive result visualization and high-throughput capabilities[61]. dPCR's advantageous attributes include sensitive detection by partitioning samples into multiple reactions, bolstering stability against inhibitory substances, and precisely quantifying even the slightest variances in target molecule counts (**Fig. 2b**).

Unlike qPCR, which relies on reference standards and is susceptible to sample quality and amplification efficiency variations, dPCR quantifies nucleic acids independently of external factors using the Poisson statistical model. In qPCR, the fluorescence of the amplified target molecules is monitored after each amplification cycle. The copy number of target molecules is quantified by mapping the relation between the cycle threshold ($Ct$) of the fluorescence amplification curve with a standard curve generated from a reference material. The amplification curves are usually vulnerable to inadequate amplification efficiency and sample quality, inhibitors, and system error. Moreover, the "analog" quantification provided by qPCR has suffered from limited inter-experiment and inter-laboratory reproducibility, as well as notable amplification bias resulting from competition of a large pool of DNA in the sample. In contrast, dPCR divides the PCR reaction mixture into tens of thousands or even millions of reaction units, and the fluorescence of each unit is detected at the end of amplification. The copy number of target molecules is calculated by the fraction of the reaction units containing the target molecules, according to the Poisson distribution. Being capable of detecting 1 mutant in 200,000 background wild-type genes, dPCR provides a 2,000 times higher sensitivity compared to qPCR (1 in 100)[62]. Consequently, dPCR is less sensitive to inhibitors than qPCR, and copy number calculation does not rely on reference materials. Compared with microplate- and chamber-based dPCR with 96 sub-reactors (in μL range) and 765 sub-reactors (in nL range), respectively, conducting dPCR in pico-liter droplet reactors naturally expands the number of partitions to at least 20,000[63], consequently enhancing the detection precision, accuracy, and dynamic range. Therefore, dPCR is eligible to provide more reliable and accurate quantification results than qPCR. For instance, Hindson et al compared the ability of dPCR and qPCR to quantify miRNAs, and the results indicated that dPCR indeed offered more accurate results[49]. However, the transition from qPCR to dPCR entails navigating the complexities of high initial costs, intricate workflows, and limited multiplexing capacities, alongside the demands of nuanced data analysis. The partitioning strategy, while instrumental for absolute quantification, might prolong operational timelines and introduce analytical complexities[64]. Such transitions embody the trade-offs between broader accessibility and enhanced precision, with dPCR trading general accessibility for exquisite precision, conditional on specialized equipment.

dLAMP complements these advancements by streamlining molecular diagnostics. It operates under isothermal conditions, removing the need for thermal cycling present in dPCR, thus expediting the analysis and simplifying the instrumentation required for nucleic acid amplification[65]. dLAMP's methodical efficiency has shown efficacy in detecting low-copy infectious agents in point-of-care scenarios, extending the reach of molecular diagnostics[66]. dLAMP's distinction from dPCR becomes prominent in its ease of integration into portable platforms, its tolerance to complex biological samples, and the ability to generate results in significantly less time. An example of this technology's prowess was demonstrated to reveal dLAMP's proficiency for rapid, accurate COVID-19 testing[67]. While dPCR remains the gold standard for sensitivity and quantification, dLAMP's speed and lower barrier to entry offer substantial benefits for scenarios where immediate results are paramount.

The cornerstone of quantitative analysis in dNAAT is the probability function, denoted as $\Pr(X = k)$. It predicts the likelihood that any given microreactor contains $k$ copies of the target gene, given $\lambda$, the mean number of target copies per microreactor:

$$f(k, \lambda) = \Pr(X = k) = \frac{\lambda^k e^{-\lambda}}{k!} \tag{1}$$

here

$k$ represents the number of occurrences ($k$ can take values 0, 1, 2, ...).

$e$ is Euler's number ($e = 2.71828…$).

! is the factorial function.

For a robust estimation of concentration, the parameters must account for a substantial number of microreactors. The probability of a microreactor having zero copies of the target, $\Pr(X = 0)$, can be precisely estimated by $e^{-\lambda}$, particularly when there is a large number of microreactors considered. From this, the concentration of the target gene copies per volume unit is determined using:

$$Concentration = -\ln\left(\frac{N_{negative}}{N_{total}}\right)/V_{microreactor} \tag{2}$$

Here $\lambda$ signifies the average number of copies per microreactor, $N_{negative}$ is the number of non-reactive microreactors, $N_{total}$ refers to the total number of microreactors, and $V_{microreactor}$ represents the volume of one microreactor.

When the template concentration is elevated, ignoring $P_r(X = 2)$ would lead to inaccuracies. Thus, mathematical corrections are necessary, which accounts for the increasing likelihood of multiple templates in a single partition.

$$\lambda' = -ln\,(Pr(X = 0)) \quad (3)$$

In this context, $\lambda'$ is an adjusted value representing the mean number of target molecules per partition, inclusive of instances featuring more than one template molecule.

The total number of target molecules denoted as T, integrates the probability of housing 1, 2, or more templates per partition, often calibrated empirically:

$$T = N_{total} \times (\lambda' + K) \quad (4)$$

Here $K$ is a correction component that compensates for the presence of multiple target molecules within a partition. Thus, $K$ is derived using the probabilities of 0, 1, or 2 target molecules detected within a partition:

$$K = p_1 \times \Pr(X = 1) + p_2 \times \Pr(X = 2) + \cdots + p_n \times \Pr(X = n) \quad (5)$$

$p_n$ indicates the prevalence of target molecules in partitions with precisely $n$ molecules, while $\Pr(X = n)$ is the probability of exact target molecules in a partition, based on the observed positive partitions.

Once T is calculated, the corrected concentration of the target in the original sample can be determined by:

$$C' = \frac{T}{V \times D} \quad (6)$$

Here $C'$ designates the corrected concentration of targets per microliter, $V$ refers to the total volume of the reaction mixture (in microliters) before partitioning, $D$ is the dilution factor if the sample was diluted before partitioning.

**Implementations of dNAAT**

The dNAAT technology has unarguably set a benchmark in life sciences and biology for its unparalleled sensitivity and precision. This technology has revolutionized environmental sampling by enabling the detection and quantification of pathogens and genetic material within complex matrices, thus serving as a critical tool for ecological assessments and biosurveillance. In the field of microbiology, dNAAT has been instrumental in dissecting microbial communities and identifying antimicrobial resistance markers with unparalleled fidelity. The oncology sector has leveraged dNAAT for the meticulous analysis of circulating tumor DNA, facilitating early cancer detection and real-time monitoring of treatment responses[68,69]. Furthermore, the tailored precision of dNAAT has been pivotal in precision medicine, where it aids in the accurate quantification of patient-specific mutations, thereby informing customized therapeutic strategies. The growing traction of dNAAT across genomics[50,70], diagnostics[20,71], and personalized medicine[19], underscores an escalating demand for advancements that are rapid, accurate, and user-friendly.

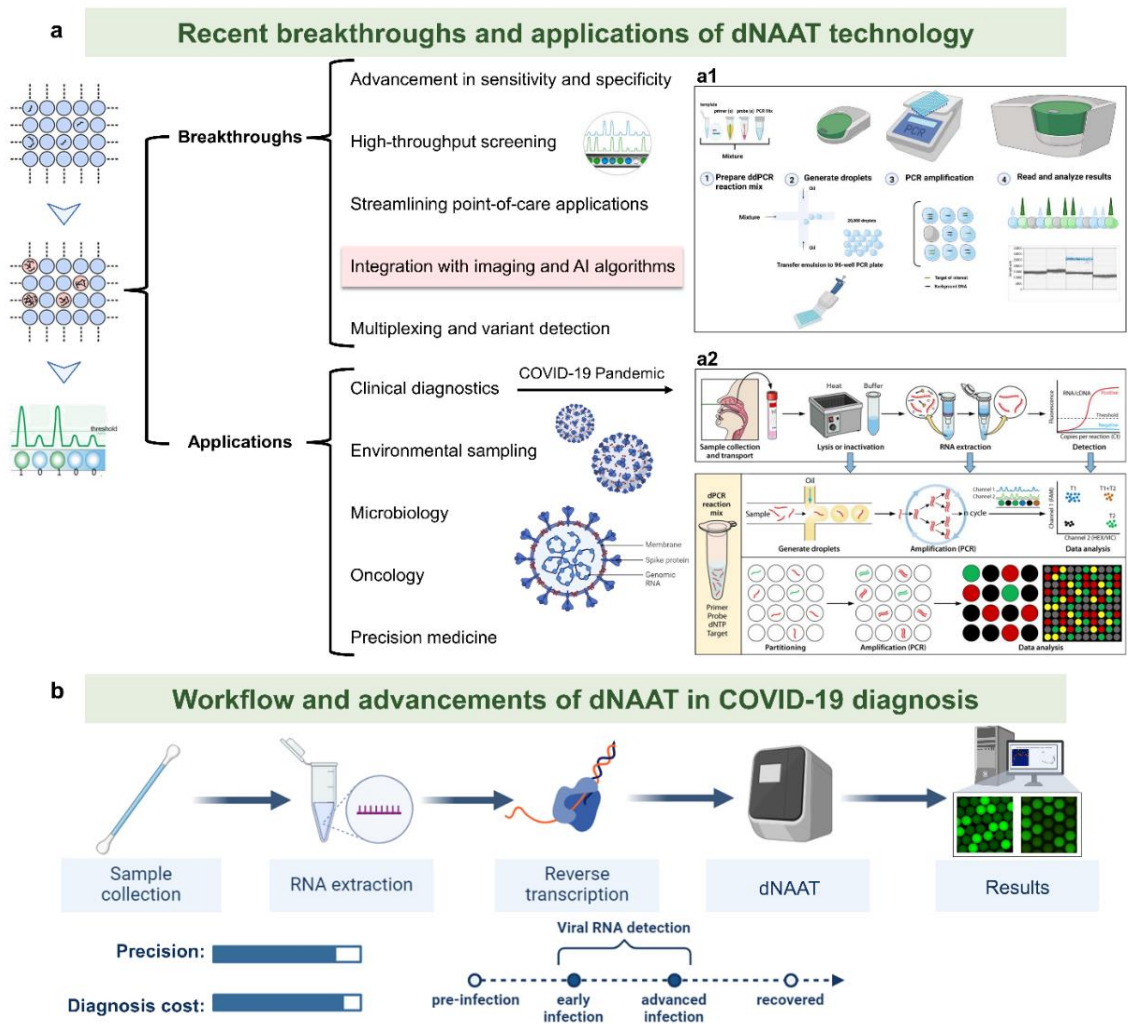

**Fig. 3 Recent breakthroughs and applications of dNAAT technology. a.** Schematic illustrating dNAAT advancements categorized into 'Breakthroughs' and 'Applications'. Breakthroughs focus on diagnostic enhancements, high-throughput capabilities, AI integration, and roles in clinical diagnostics and precision medicine. It highlights sensitivity, specificity, screening efficiency, point-of-care innovations, AI-assisted data analysis, and advanced variant detection. Applications emphasize dPCR's use in clinical diagnostics, including COVID-19 testing, environmental assessments, microbiology, oncology, and personalized medicine, underscoring its significant contribution to pandemic response efforts. **a1.** Schematic of the ddPCR process: loading ddPCR mix and DNA sample into a multi-channel cartridge, droplet generation under negative pressure, individual amplification of water-in-oil droplets, and analysis of each droplet post-PCR amplification[46]. © 2022 MDPI **a2.** Workflow for SARS-CoV-2 detection via RT-dPCR: from sample collection and potential detection points (crude lysates, purified RNA, or post-RT-qPCR) to the ddPCR and chip/chamber-based dPCR processes[72]. © 2022 American Society for Microbiology. **b.** Workflow and technological advancements of dPCR in COVID-19 diagnosis: sequential process from sample collection, RNA extraction, reverse transcription, and viral RNA detection using dNAAT, culminating in results display. It outlines infection progression across five stages: pre-infection, early infection, peak infection, declining infection, and recovery. Additionally, it visually represents the precision and cost-effectiveness of diagnosis at various infection stages, emphasizing the critical role of timely and accurate testing.

Recent breakthroughs in dNAAT technology have heralded a new era characterized by enhanced sensitivity and specificity, thereby augmenting the detection of ultra-low abundance targets[73]. **Fig. 3a** provides an overview of the significant advancements in dNAAT technology, divided into 'Breakthroughs' and 'Applications'. 'Breakthroughs' detail enhancements in sensitivity and specificity, high-throughput screening capabilities, innovations in point-of-care applications, improved integration with imaging and AI for in-depth data analysis, and refined techniques for multiplexing and variant detection. The 'Applications' section showcases the deployment of dNAAT in fields including clinical diagnostics. This emphasizes its role in COVID-19 testing, environmental sampling, microbiology, oncology, and precision medicine, signposting its impact on the pandemic through diagnostic workflows. The advent of ddPCR has showcased the technology's prowess in detecting variants with a sensitivity threshold as low as 0.1%, a capability that is crucial for early disease diagnosis, minimal residual disease monitoring, and mutation detection[74]. The development of high-throughput dPCR systems has underscored the technology's advancements in molecular diagnostic standardization, enabling the simultaneous processing of thousands of samples without compromising the intricate sensitivity and specificity required, even in complex biological matrices[75]. Furthermore, the integration of real-time analysis within dPCR platforms represents a transformative development, facilitating dynamic monitoring and immediate interpretation of reactions[76]. The incorporation of AI-driven algorithms into dPCR workflows has significantly enhanced predictive power and automation, propelling the fields of precision medicine and genomic research forward. Additionally, advancements in multiplexing and variant detection technologies have expanded the scope of dPCR's application in complex genomic investigations. dLAMP distinguishes itself from dPCR by eliminating the need for thermal cycling and thereby simplifying the instrumentation required. Moreover, dLAMP's ability to amplify DNA with high specificity and rapidity, even in the presence of non-target sequences, offers a significant advantage in the detection of infectious diseases and genetic mutations. Despite these benefits, dLAMP's sensitivity and quantitative accuracy are areas where dPCR still holds superiority, especially in applications requiring the detection of low-abundance targets and precise quantification for clinical decision-making. The continuous refinement and integration of these technologies will undoubtedly enhance our capabilities in disease diagnosis, environmental monitoring, and personalized medicine, marking a significant stride toward the advancement of molecular diagnostics.

## 3. dNAAT under the context of the COVID-19 pandemic

The utility of dNAAT, highlighted by the widespread adoption of dPCR and dLAMP, has been particularly emphasized during the COVID-19 pandemic. These methodologies offer high sensitivity, precision, and the capability for absolute quantification of viral RNA[77], surpassing traditional RT-PCR in several aspects. Notably, traditional reverse transcription-quantitative polymerase chain reaction (RT-qPCR) has demonstrated limitations in viral load analysis, which is crucial for evaluating disease severity and transmission dynamics[78]. Conversely, dPCR enhances detection limits, serving as a critical instrument for COVID-19 management, including establishing discharge criteria based on viral load quantitation[79,80]. This accuracy is paramount for the proper diagnosis and surveillance of SARS-CoV-2, thereby providing a pivotal tool for accurate diagnosis and monitoring of the virus' spread.

The dPCR streamline, depicted in **Fig. 3b**, encompasses a series of steps from sample collection, covering RNA extraction, reverse transcription, digital PCR, and concluding with result analysis. This process is adept at diagnosing both early and advanced stages of infection. Unlike conventional PCR, dPCR provides absolute quantification of viral load, offering more accurate assessments of infection levels[81,82]. This precise quantification enables the detection of low-level RNA found in asymptomatic and pre-symptomatic cases[83,84]. This capability is instrumental in monitoring the progression of the disease and the effectiveness of treatment strategies[85]. Furthermore, dPCR excels in standardizing viral load measurements across different laboratories, improving the consistency and reliability of data[86]. dPCR has shown superiority in detecting variants[24] and identifying specific mutations within the viral genome[87,88], thereby facilitating the tracking of emergent SARS-CoV-2 variants[70].

Environmental surveillance, including wastewater monitoring, exemplifies another vital application of dPCR, proving its efficacy in tracing low levels of SARS-CoV-2 RNA[89,90]. RT-dPCR has demonstrated suitability for low-level detection in environmental samples, providing early indicators of community infection levels and informing public health interventions[91,92]. Such applications underscore dPCR's role beyond clinical diagnostics, including drug development and longitudinal studies. It has been used to quantify viral loads in investigations of household transmission dynamics, highlighting variations in viral load and transmissibility among different variants of concern[93,94]. By reducing false negatives and improving reliability, dPCR has overcome limitations inherent in conventional methods. For instance, ddPCR detected SARS-CoV-2 in previously negative RT-PCR cases, showcasing its potential to reduce false-negative rates[95].

Transitioning to dLAMP, this methodology introduces efficiency through isothermal amplification, contrasting dPCR's thermocycling requirements[45]. This difference is paramount in contexts necessitating rapid diagnostic results, such as the COVID-19 pandemic, where rapid and reliable diagnostic methods are crucial for timely isolation and intervention[67]. dLAMP's rapidity and compatibility with point-of-care settings underscore its viability where traditional laboratory infrastructure is absent. Researchers have demonstrated dLAMP's capacity to directly detect SARS-CoV-2 RNA from patient samples sans RNA extraction, marking a significant stride towards expediting the diagnostic process[96]. Consequently, dLAMP seamlessly caters to the surge in diagnostic demands, delivering rapid, reliable testing even to the most resource-constrained regions[97]. Bridging the rapid diagnostic needs with decentralized testing capabilities, dLAMP amplifies the reach and efficiency of molecular diagnostics in combatting COVID-19.

## 4. Comparative analysis and shared features of commercial dPCR systems

Commercial dPCR systems have transformed nucleic acid quantification by enabling highly sensitive and accurate measurements. **Fig. 4a** and **Table 1** present a comparative assessment of available commercial dPCR systems, offering researchers insight into the SOTA instrumentation fueling the diagnostic evolution. These systems utilize microfluidics or droplet technologies to partition a sample into numerous micro-reactions, enabling the detection of rare mutations and precise quantification of DNA copy numbers, even at very low concentrations[98].

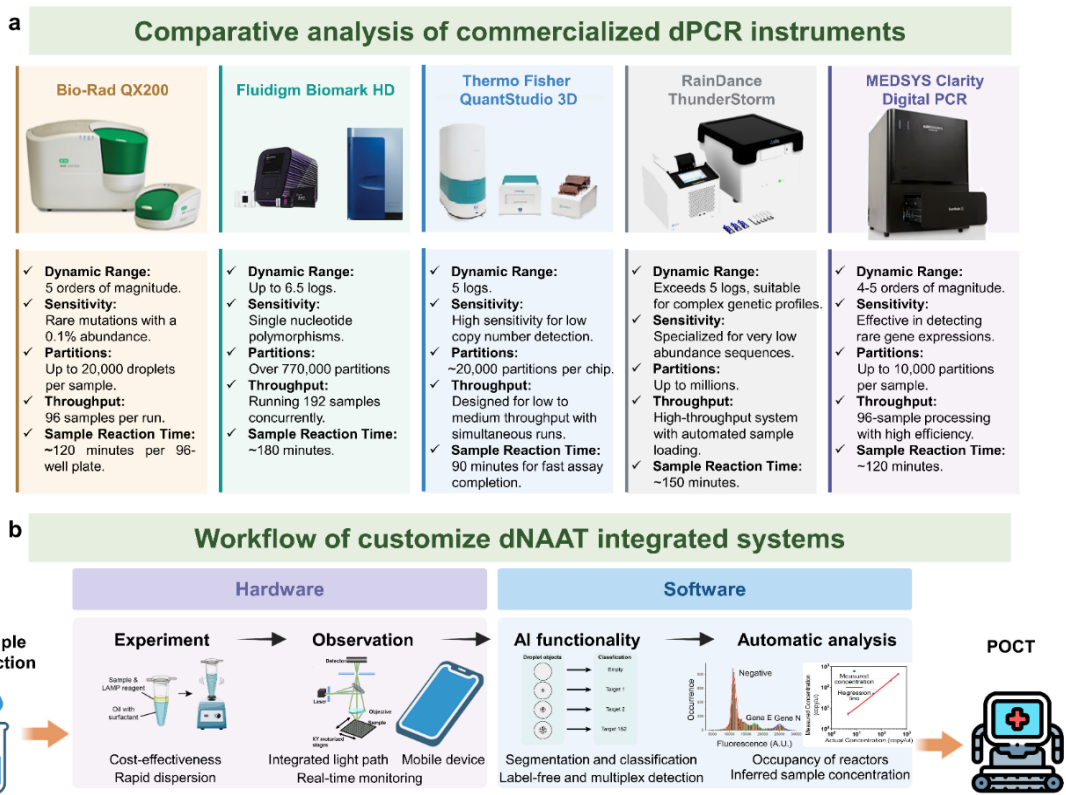

**Fig. 4 Comparative analysis of commercialized dNAAT instruments and overview of customized dNAAT platforms. a.** Comparative analysis of leading commercial dPCR instruments: Bio-Rad QX200, Fluidigm Biomark HD, Thermo Fisher QuantStudio 3D, RainDance ThunderStorm, and MEDSYS Clarity Digital PCR. This panel highlights each instrument's dynamic range, sensitivity, partitioning capabilities, and sample reaction time, along with throughput capacity and the number of samples processed per run. It provides insights into the unique strengths and specifications of each platform, aiding in their suitability assessment for various diagnostic needs. **b.** Workflow of customized dNAAT integrated systems: from sample collection and experimentation to observation and point-of-care testing (POCT). The workflow details the system's cost-effectiveness, integrated light path, real-time monitoring capabilities, mobile device connectivity, and AI-driven segmentation and classification. It emphasizes the system's ability to perform label-free and multiplex detection, offering a glimpse into the future of streamlined, efficient, and accurate molecular diagnostics. Each panel provides a comprehensive overview of both commercial and customized dNAAT platforms, highlighting their capabilities and potential impacts on molecular diagnostics.

Bio-Rad's QX200 and Thermo Fisher's QuantStudio 3D set the industry standard with a dynamic range of 5 logs and the sensitivity to detect target frequencies as low as 0.1%. The

Fluidigm Biomark HD extends the dynamic range to an impressive 6.5 logs and boasts the partitioning of over 770,000 reactions, providing unmatched sensitivity over a broad spectrum of DNA concentrations. Meanwhile, the RainDance ThunderStorm system highlights customizable partitioning that can reach into the millions, offering granular analysis capabilities for complex genetic inquiries.

Throughput capacity is another determining factor in selecting a dPCR system. The Bio-Rad QX200 and JN MEDSYS Clarity Digital PCR cater to standard laboratory throughput with the capability of processing up to 96 samples per run. The Fluidigm Biomark HD is suitable for high-throughput research environments, offering double the capacity at 192 samples per run. The partition number further augments a system's utility. The extraordinary partition count made available by the Fluidigm Biomark HD contrasts with the Bio-Rad and Thermo Fisher systems, each offering partitions in the range of approximately 20,000. The systems also diverge when it comes to running time. The Thermo Fisher QuantStudio 3D leads with the fastest turnaround, completing cycles in 90 minutes, ideal for time-sensitive research. Both the Bio-Rad QX200 and JN MEDSYS Clarity Digital PCR follow closely, with assays finalizing in approximately 120 minutes, delivering a balance of efficiency and thoroughness. In high-throughput setups requiring comprehensive analysis, the Fluidigm Biomark HD and RainDance ThunderStorm necessitate longer processing times.

Despite of these matured commercialized dPCR machines, early-stage dPCR platforms are also assessed. **Table 2** provides a comparative evaluation of the prototype platforms from leading manufacturers such as QIAGEN, Life Technologies, SlipChip, Sysmex Inostics, Stilla Technologies, Rubicon Genomics, and Leiden University Medical Center. It outlines key features and performance metrics, including the platform and function of each system, the mode of detection (end-point or real-time), the number of partitions ranging from 8,500 to 20,000 for sample segmentation, the range of detection channels available for multiplexing, the time required per run from approximately 1 hour to several hours, and the degree of automation integrated into the system. Additionally, the table highlights the primary applications and use cases for each platform, ranging from rare mutation detection to nucleic acid quantification and infectious disease diagnostics. This comprehensive overview assists researchers and laboratory technicians in selecting a dPCR platform that aligns with their specific research goals and operational needs, emphasizing factors such as sensitivity, throughput, and cost-effectiveness.

Notwithstanding their technical prowess, all dPCR systems embrace a core set of features essential for their contribution to modern molecular diagnostics: The commonality of these systems lies in their partitioning strategy, ability to perform absolute quantification, multiplexing capability, and sensitivity. (1) Increased sensitivity: dPCR systems can detect and quantify rare nucleic acid targets with high precision and sensitivity[126]. They can accurately determine the absolute number of target molecules present in a sample, even when they are present in low abundance. (2) Improved accuracy and reproducibility: dPCR systems eliminate the need for standard curves, making it less prone to variations caused by PCR efficiency[86]. This results in improved accuracy and reproducibility of the measurements. (3) Absolute quantification: Unlike traditional PCR systems, dPCR systems provide absolute quantification of target molecules, as they counts the number of target molecules directly. This eliminates the need for external calibration curves and improves the accuracy of quantification. (4) Increased dynamic range: dPCR systems have a wider dynamic range compared to traditional PCR methods. They can accurately detect and quantify both high and low-copy number targets in the same reaction, without the need for dilution or concentration steps [127]. (5) Reduced reliance on reference genes: dPCR systems allow for the quantification of target genes without the need for reference genes [64]. This is particularly advantageous in situations where suitable reference genes are not available or are not stably expressed.

However, practical considerations such as budget constraints, workflow complexity, and multiplexing capacity pose unique challenges. To be more specific, the challenges are: (1) Cost: Commercial dPCR systems can be expensive, both in terms of the initial instrument purchase (more than 70,000 USD) and the ongoing costs of consumables and reagents. This can limit their accessibility to some researchers or laboratories. (2) Workflow complexity: dPCR systems often require more complex workflows compared to traditional PCR methods. This may involve additional steps such as partitioning the sample into individual reactions or using specialized consumables. Users need to be familiar with these workflows and ensure proper training and optimization. (3) Limited multiplexing capacity: Some commercial dPCR systems have limited multiplexing capacity, meaning they can only detect and quantify a limited number of targets in a single reaction. This may be a limitation for researchers who require simultaneous detection of multiple targets. (4) Time-consuming: dPCR can be more time-consuming compared to traditional PCR methods. The partitioning of the sample and the analysis of individual reactions can increase the overall assay time. (5) Data analysis challenges: dPCR generates a large amount of data, which can be challenging to analyze and interpret. Specialized software and bioinformatics expertise may be required to analyze the data accurately. Even the drawbacks are not inherent to all commercial dPCR systems, but rather general considerations that researchers should be aware of when choosing and utilizing these technologies. The GeneXpert system, pioneering a cartridge-based integrated dPCR approach, exemplifies innovation yet acknowledges the trade-offs, inherently challenging in terms of multiplexing due to a single chamber. In conclusion, while commercial dPCR systems vary in their dynamic range, sensitivity, and sample throughput, their selection should be considered within the scope of the laboratory's specific research goals, emphasizing cost-effectiveness and operational efficiency.

Customized dNAAT systems have demonstrated remarkable flexibility, integration, and adaptability, addressing the continuously evolving demands of COVID-19 diagnostics and beyond. **Fig. 4b** illustrates the workflow and innovation of SOTA integrated dNAAT systems, tracing the path from sample collection and experimentation to observation and POCT. The illustration delineates the advancement of dNAAT systems into two core components: 'hardware' and 'software.' Each element is scrutinized for breakthroughs, with 'hardware' improvements evaluated through enhancements in experimental setups and observational capabilities, and 'software' advancements gauged by AI-driven functionality and automated data analysis. The workflow accentuates the system's proficiency in conducting label-free and multiplex detection, providing a preview of the future landscape of molecular diagnostics which is characterized by its streamlined processes, efficiency, and precision.

## 5. The combination of AI and dNAAT for enhanced diagnostic precision

Although dNAAT technology has emerged as a cornerstone in the realm of clinical diagnostics, its broader application, specifically within primary healthcare and epidemic-stricken regions, is stymied by a spectrum of operational challenges. These encompass the enhancement of sample throughput and automation, a reduction in procedural complexity and associated expenditures, and the establishment of universal clinical laboratory standards. Conventional detection modalities such as commercialized machines, flow cytometry, and manual fluorescence imaging, expose pronounced limitations. Commercialized dPCR platforms such as those from Bio-Rad and Raindance, while precise, pose issues with cost, complexity, limited multiplexing, time demands, and data analysis. High initial and recurring expenses may deter use, especially in budget-strapped labs [26,27]. While flow cytometry stands as a revered 'gold standard' for cell analysis and sorting, it is less practical for the high-throughput needs of dPCR due to its high costs, the necessity for expert operation, and complex sample preparation[29]. Similarly, fluorescence imaging, reliant on manual thresholding via software such as Image J[128], Fiji[70], and CellProfiler[30], are ensnared by labor intensity and lack scalability[129,130]. Furthermore, there's an eminent challenge of forging a universal gold standard for data classification. Prevailing paradigms furnish a circumscriptive informational palette and habitually obfuscate nuances, including microreactor uniformity[131,132] and fluorescence intensity[133]. Predominantly used commercial applications (e.g., QuantaSoft, Bio-Rad) resort to manual threshold calibrations (employing K-means or KNN data clustering method)[98], thus embedding user bias and potential data unreliability into the process.

This is where deep learning, an AI subset, exerts prominence, adeptly navigating the convolutions within biomedical sensory data. Brandishing an astute acumen for feature classification, anomaly detection, and multiplex analysis, deep learning eclipses conventional techniques[134]. Its adaptability to fluctuations in nucleic acid amplification data begets a refinement in dPCR workflows, attenuating the accustomed manual oversight. This pivot is accentuated by neural networks' end-to-end learning finesse, fostering autonomous precision refinement concurrent with dPCR datasets' evolving intricacies. AI's momentous aptitude in adapting to nucleic acid amplification variance fortifies this evolution, mitigating the onus traditionally besieging conventional approaches. This autonomous learning cadence, underscored by potent backpropagation and non-linear representational capabilities, elevates neural networks as instrumental for expeditious knowledge distillation from compendious data troves[135]. Herein, AI emerges as a transformative linchpin, augmenting dNAAT applications as illustrated in **Table 3**. This confluence begets marked improvements in system efficiency[37,38,41,136–138], automation[39,136–140], adeptness[37–39], precision[37,137,138].

The prevalence of traditional machine learning techniques, such as automated ovo-SVM systems and supervised k-NN strategies, has been pivotal steering dPCR and dLAMP towards rigorous detection and classification. Exemplified by automated ovo-SVM systems wielding support vector machines for class segregation[140] and k-NN methods employing labeled training data, these modalities constitute quintessential assets for automated detection and classification, bolstering dPCR's analytical ambit[136].

Amidst this matrix, the ascension of advanced deep learning motifs has broadened the horizons for complex feature discernment within dPCR and dLAMP paradigms. Groundbreaking models like Deep-qGFP[39], SCAD[41] and StratoLAMP[40], amalgamate supervised deep learning with innovative network architectures, delivering precision and automated quantification of microreactors. Employing Yolov5 and an RPN (Region proposal networks) modes, Deep-qGFP offers precise and automated quantification of green fluorescence profiler-labeled microreactors with an accuracy higher than 96.23%. In SCAD, a supervised deep learning model based on ViT model and tSNE has been successfully demonstrated for multiplex dPCR detection in a single fluorescent channel, which is not reachable with manual differentiation modality. StratoLAMP heralds a pivotal advancement in dLAMP by introducing label-free and multiplex detection capabilities through visual stratification of precipitate byproducts, thus obviating the need for fluorescence-based indicators. This innovation promises to transform the accessibility and efficiency of dNAAT technologies, particularly in resource-limited settings and for broad-based pathogen surveillance. This surge encompasses hybrid methodologies that amalgamate various techniques such as the fusion of Mask R-CNN with the Gaussian Mixture Model (GMM)[37], fostering enhanced performance in nucleic acid quantification. Combining Attention DeepLabV3+ with the circle Hough transform (CHT), droplet detection and classification are achieved with an accuracy higher than 97%, even in low-quality dPCR images[139]. These multifaceted learning approaches, juxtaposed with experimentation, usher in a novel epoch of dNAAT image analysis, one characterized by precision, robustness, and an economy of scale[137].

Advancements in computer science have led to the development of high-performance models for object detection. Researchers have been actively working on improving existing models and creating novel variants to expand the capabilities of this field. An enhanced YOLOv5 algorithm with the CBAM attention mechanism has been developed to accurately identify negative and positive droplets, even in complex backgrounds. This model is characterized by its speed, adaptability, and suitability for deployment on mobile or cloud platforms, with minimal error and missing rates (0.65% and 2.17%, respectively)[138]. Additionally, a few-shot learning technique combining an improved YOLOv3 model with RBTM and STAM has significantly improved detection accuracy, speed, and cost-effectiveness, particularly in scenarios with limited data and challenges. This approach reduces labeling time by 70% while achieving an impressive accuracy of 98.98%[38]. These advancements hold great promise for enhancing dPCR image analysis, enabling more efficient and accurate research in the field.

To encapsulate, advancements in cutting-edge AI modalities have unfurled thus: (1) High automation level: The process is highly automated, enhancing efficiency and reliability. (2) Broad applicability: A generalist approach caters to a wide array of research demands and experimental settings. (3) Increased accuracy: Improves sensitivity and precision, contributing to higher quality outcomes. (4) Processing efficiency and real-time capability: Time-efficient and user-friendly, reducing the need for repetitive testing and additional confirmatory tests, thereby improving cost-effectiveness. Exceptional speed in absolute quantification and processing large datasets enables real-time analysis. (5) Data efficiency: Capable of yielding meaningful insights from limited datasets without the need for extensive data collection and training. (6) Personalized diagnosis: The incorporation of AI in dNAAT data analysis advances personalized medicine by providing tailored insights. (7) Multiplexing capabilities: Enables comprehensive molecular diagnostics and quantitative analysis through multiplexing. (8) Public health impact: Supports informed public health decisions.

Meanwhile, the weaknesses can be classified as: (1) Model interpretability: Issues with model transparency and interpretability may obscure decision-making processes. (2) Data quality dependency: The model relies on high-quality data, with a risk of overfitting, highlighting the importance of data integrity. (3) Scalability issues: Encounters scalability challenges in managing larger datasets or very limited data on highly variable samples or conditions. (4) Generalization difficulty: Primarily designed for certain molecular techniques or applications, limiting broader utility. (5) Accessibility limits: The model may present challenges in accessibility and usability for users without specialized knowledge. (6) Cross-reactivity risk: Multiplex assays may encounter cross-reactivity, affecting specificity. (7) Data privacy concerns: Shared or cloud platform use raises questions about data privacy and security. (8) Color differentiation limits: Bound by color-based differentiation, restricting alternative multiplexing strategies.

The identified strengths can be categorized into two distinct groups: Strengths 1-4 (labeled with *) represent universal advantages, showcasing benefits widely recognized in the field. In contrast, Strengths 5-8 (labeled with #) are unique advantages that stem from bespoke designs, reflecting the innovative and customized nature of the approaches. On the downside, Weaknesses 1-5 (*) are general challenges not uncommon across the algorithms, illustrating areas where broad improvements can be beneficial. However, Weaknesses 6-8 (#) are specific to the works, highlighting targeted areas for refinement that directly relate to distinct methodology and system configuration. These personalized challenges underscore the need for tailored solutions to enhance system performance and reliability.

The integration of AI with dNAAT heralds a transformative change in diagnostics, propelling us into a new era of high-throughput automation and sophisticated image analysis that reshapes conventional methodologies. The employment of AI techniques ranging from attention mechanisms and gradient-based approaches for enhanced model clarity[37,41,136], to data preprocessing for improved quality[41,136,139], and the adoption of transfer learning for increased adaptability across different domains[38,39,139,140], fortifies the position of AI as a leading force in advancing the dNAAT arena. Moreover, the implementation of privacy-conscious federated learning further strengthens this advancement[137,138]. AI's incorporation not only broadens access but also addresses the challenges associated with colorimetric assessments and enforces thorough optimization procedures. This coalescence directs the field of molecular diagnostics toward unparalleled precision and operational excellence.

## 6. Integrated dPCR systems for automatic and high-throughput nucleic acid quantification

Building on the elevated diagnostic precision afforded by AI in digital NAAT image analysis, the crucial next step lies in transcending the cumbersome mentalities of traditional NAAT-based molecular diagnostics. Conventional diagnostics necessitate a multi-step, labor-intensive gamut, from sample acquisition to complex PCR cycles, that demands specialist handling and risks inadvertent contamination or laboratory-acquired infections. The axiom of modern-day diagnostics is a seamless transition from cumbersome laboratory procedures to a proficient, integrated molecular diagnostic system on a chip design. This advancement coalesces DNA/RNA extraction and amplification processes into a unified device, championing the cause for in-home diagnostics and embodying the epitome of personalized, preventive medicine. The envisioned system streamlines NA extraction from biological specimens via fluidic devices and couples it with an integrated amplification unit, proffering a robust solution that negates the need for external, elaborate machinery.

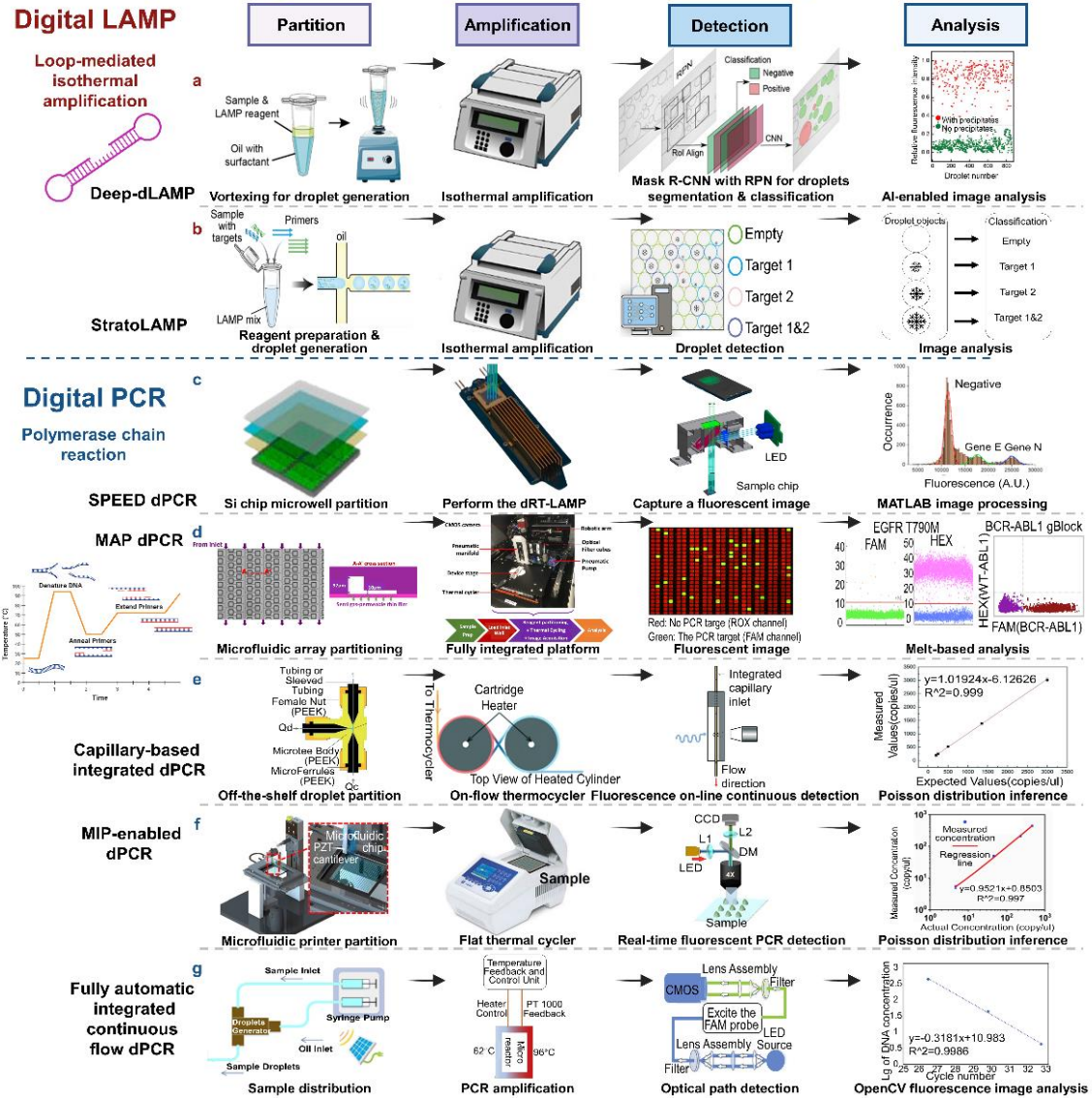

**Fig. 5 Advanced workflows in dLAMP and dPCR systems for enhanced automating precision and throughput. a.** Deep-dLAMP employs off-the-shelf laboratory hardware and an AI model based on Mask R-CNN for label-free, cost-effective nucleic acid quantification, using a standard vortex mixer for polydisperse emulsion generation. Adapted from ref.[36], CC BY 4.0. **b.** StratoLAMP, an AI-powered dLAMP system, enables label-free and multiplex detection through visual stratification of precipitates, advancing label-free detection methodologies. Adapted with permission from the ref.[40]. © 2024 National Academy of Sciences **c.** SPEED dPCR is a smartphone-integrated, handheld system with accuracy comparable to conventional dPCR machines. Its compact and lightweight design includes miniaturized modules for thermal cycling, image capture, and wireless data transmission, using a silicon-based dPCR chip. Adapted with permission from the ref.[99]. © 2023 Elsevier B.V. **d.** MAP dPCR system streamlines workflows and minimizes contamination risks with a micro-molded plastic microfluidic consumable. It integrates pneumatic sample loading, thermal cycling, and tri-color fluorescence image acquisition, controlled by comprehensive software for real-time and melt-based partition analysis. Adapted with permission from the ref.[142]. © 2019 Springer Nature Limited. **e.** Capillary-based integrated dPCR system utilizes an HPLC

T-junction for droplet generation and a long HPLC capillary connecting the droplet generator to a capillary-based thermocycler and cytometer, ensuring zero droplet loss and no cross-contamination. Temperature control is managed by an ARDUINO UNO REV3 with PID control and thermocouples. Adapted with permission from the ref.[29]. © 2018 Royal Society of Chemistry. **f.** MIP-enabled ddPCR platform uses microfluidic impact printing technology for sample partitioning, achieving over 95% efficiency in compartmentalization and a sample-to-answer throughput in less than 90 minutes. Adapted from ref. 143. © 2020 Elsevier B.V. Adapted with permission from the ref.[143]. © 2020 Elsevier B.V. **g.** Fully automatic integrated continuous flow dPCR system offers one-step operation with complete software integration for control and data processing, incorporating continuous-flow principles. Adapted with permission from the ref.[144]. © 2020 Elsevier B.V. Each panel illustrates the progression from partitioning through amplification to detection and analysis, showcasing how advanced dLAMP and dPCR systems enhance precision and practicality in molecular diagnostics.

Despite the persistent strides towards POCT adaptability, achieving a fully integrated sample preparation from unrefined biologic samplings for NA-based molecular diagnostics has been a formidable challenge. **Fig. 5** and **Table 4** delineate the comprehensive workflow and various integrated dNAAT systems that have been developed to enhance the automation and throughput of nucleic acid quantification. This workflow is dichotomized into 'hardware' and 'software' realms, each undergoing rigorous advancement: pragmatic 'hardware' developments are gauged through experimentation and observational evolutions, whereas 'software' progressions are calibrated through AI integration and data analysis automation. The dPCR and dLAMP technologies are illustrated respectively, sketching the continuum from sample partitioning to analytic denouement with a range of significant strengths including automation[29,36,99,145], cost-effectiveness[40,146–148], high-throughput[29,142,144,145], POCT[44,51,148–150], minimized risks of cross-contamination[25,51,142,145], adaptability to diverse samples[145,146], and improved reproducibility[151]. These avant-garde systems have been meticulously honed at each juncture to amplify precision and functionality, as well as epidemiological tracking[148] and control of infectious diseases.

The advancements in reagent partitioning can be classified into improved precision, decreased cross-contamination and sample loss through automation or all-in-one platforms, reduced cost for microfluidic set-up, and special functions such as rapid nucleic acid extraction[150] and self-priming. An integrated lab-on-a-disc (LOAD) device realized all ddPCR processes performed inside one whole chamber for quantitively screening infectious disease agents[51]. The microfluidic array partitioning (MAP) dPCR reduces cross-contamination using off-the-shelf reagents and micro-molded partitions[142]. Droplet dPCR through microfluidic impact printing (MIP-dPCR) achieves precise partitioning with a disposable chip, offering automation and high throughput[143]. SPEED dPCR supports various dPCR uses with consumable silicon chip partitions but faces limitations in cross-contamination risk and scalability/cost challenges[99]. A smartphone-based droplet dLAMP device that integrated immiscible phase filtration, reagent mixing, microdroplet generation, and droplet tiling was developed for detecting low abundance cfDNA and EGFR L858R mutation with high sensitivity in 1 h[150]. Utilizing a simple and cost-effective partitioning method, SlipChip dPCR uses movement, or "slipping" of two plates and separates the wells, creating 1280 reaction compartments with preloaded oil[112,153]. This innovative approach enables efficient compartmentalization without the need for complex fabrication processes or intricate manipulation systems. Similarly, in Deep-dLAMP, the distribution using vortexing ensures uniform distribution of LAMP reagents while directly eliminating the need for a microfluidic chip[36]. Other hardware advancements include diverse pumps for actuation and precise fluid control[51,70,87,149,154].

After partitioning, the amplification speed, automation, and throughput get enhanced via advancements in hardware and software such as parallel reactions. A large amount of dNAAT reactions, both dPCR and dLAMP, were performed using lab PCR thermal cyclers[36,39,50]. A rapid ddPCR system realized ultrafast multiplexed detection of SARS-CoV-2 RNA via a 5-min ddRT-PCR process using in-situ heater arrays for efficient thermal cycling[155]. The LOAD ddPCR utilizes a transparent circulating oil-based heat exchanger for rapid heating and cooling[51]. It provides an effective way to regulate the temperature for ddPCR experiments and enables the following transmission-based fluorescence detection. An all-in-one dLAMP platform consists of a compact analyzer and a disposable microfluidic reagent compact disc and is capable of processing four samples simultaneously within a 50-minute turnaround time[156]. The device provides real-time fluorescence nucleic acid testing with automated and scalable sample preparation capability. SPEED dPCR demonstrates a smartphone-integrated handheld dPCR system with miniaturized modules for thermal cycling, image taking, and wireless data communication. The prototype is compact (100 mm x 200 mm x 35 mm) and lightweight (400g) while performing comparable accuracy with commercial dPCR machines[99]. The integrated continuous flow dPCR system offered high throughput and features like reagent refrigeration and temperature control for cost-effectiveness[144]. It supports various dPCR applications and demonstrates a strong correlation coefficient (0.9986). Despite its significant advancements, the device has limitations such as contamination risk, limited throughput in specialized setups, and the need for specialized technical expertise. In an all-in-one capillary-based dPCR system, a long HPLC capillary connects the generator with both a capillary-based thermocycler and a capillary-based cytometer[29]. Thus, cross-contamination or droplet coalesce is minimized by eliminating transferred manually to modules for amplification and detection.

During the thermal cycling, fluorescently labeled probes bind to specific sequences within the target DNA and release a fluorescence signal in positive reactions where amplification occurs. In the detection process, advancements include high-throughput screening, enhanced optical systems, multiplexing capability, integration, accessibility and cost reduction. A highly integrated real-time amplification and detection dPCR device integrates a thermocycler and a fluorescence image setup with the microfluidic chip using 3D-printed parts and off-the-shelf electronics[25]. It provides dynamic continuous observation, and better accuracy than end-point detection instruments while is still affordable. MAP dPCR systems accurately quantify ultra-rare genetic events (<1.0%) and excel in precision cancer monitoring, even at low DNA concentrations[142]. Individual partitions minimize cross-contamination and support high-throughput analysis. Capillary-based dPCR systems utilize continuous detection with excitation light and a CMOS camera for real-time fluorescence monitoring, with comparable or even better accuracy than fluorescence images[144]. Deep-dLAMP software utilizes advancements like Mask R-CNN with RPN and the SGD optimization algorithm to enhance droplet segmentation and achieves a high detection limit (5.6 copies $\mu l^{-1}$) and dynamic range (37.2 to 11,000 copies $\mu l^{-1}$)[36]. It enables automated workflows and multiplex analysis within the isothermal amplification while may have limitations in detecting low-abundance targets.

StratoLAMP offers label-free and multiplex capabilities through visual stratification of the precipitate byproduct, eliminating the requirement for fluorescence[40]. However, StratoLAMP faces challenges when different droplet diameters are intended or different primer sets are used, new models should be trained to accommodate the new experimental settings, which highlights its inability to generalize across diverse experimental conditions. These advancements revolutionize precision cancer monitoring and enable POCT in resource-limited areas.

The final step is the analysis of the fluorescence data to determine the absolute concentration of the target DNA in the original sample. There have been substantial software advancements for the analysis and interpretation of complex dPCR data. These include algorithms that are capable of handling large datasets, automating calculations, and providing intuitive visualization tools. The integrated continuous-flow dPCR offers complete control and data processing capabilities. Developed in Python, the device control software enables seamless communication and control[144]. It incorporates fluorescence image processing using OpenCV, enhancing analysis. The high correlation coefficient (0.9986) confirms the device's reliability and accuracy in quantifying DNA, making it a cost-effective solution with versatile software support. It also simplifies the testing process with a one-step operation, making it user-friendly. Deep-dLAMP and SPEED dPCR utilize advanced technology to enhance their image analysis capabilities. Deep-dLAMP benefits from software tools like Origin and ImageJ, enabling comprehensive statistical analysis and data manipulation[36]. It also uses a Mask R-CNN model with RPN for accurate target detection and segmentation. SPEED dPCR incorporates a chip mask to correct for imaging system aberrations, ensuring high-quality images[99]. Software advancements include an image-to-answer algorithm and the circle Hough transform algorithm, enabling accurate analysis and extraction of valuable information[157]. Moreover, customize Android software for smartphone-based mobile dPCR devices also enhanced automation and ease-to-use[44,99]. These advancements enhance the accuracy, reliability, and interpretability of image analysis in dNAAT processing. Allowing multiple samples to be processed simultaneously, the advancements in data analysis speed up the workflow and make dNAAT a viable option for large-scale studies.

The assessment outlines a comprehensive analysis of both strengths and limitations associated with the system in question, elaborating on how it contributes to the dynamism and potential constraints in the field of nucleic acid quantification related applications. It provides a thorough analysis of the strengths and limitations of the dPCR and dLAMP systems, highlighting its contributions and constraints in the field. The system's strengths include: (1) Full automation enhances workflow efficiency and reduces manual errors, leading to consistent and reliable results. (2) Hardware improvements bolster cost-effectiveness and performance, enabling rapid thermal cycling crucial for PCR processes. (3) Software enhancements facilitate high-throughput screening and rapid data analysis. (4) The system's versatility supports a diverse range of dPCR applications across research and diagnostic fields. (5) High throughput capabilities allow simultaneous processing of numerous samples, maximizing efficiency. (6) Multiplex analysis enables the detection of multiple targets in a single run, saving time and resources. (7) Mobile and portable systems extend the technology's reach to resource-limited areas and POCT. (8) Label-free detection simplifies assay preparation and reduces costs. (9) Design minimizes cross-contamination risks, enhancing result reliability. (10) Reduced sample and reagent volumes lower operational costs and conserve valuable samples. (11) The system's adaptability to complex samples ensures broad applicability, aiding in (12) epidemiological tracking with precise and rapid detection. (12) Improved reproducibility is crucial for consistent experimental outcomes.

However, the system also faces limitations. (1) Specialized hardware requirements may restrict real-time analysis to those with access to the necessary equipment. (2) Connectivity and software compatibility issues can hinder efficient operation and data analysis. (3) High miniaturization and open capillary systems increase cross-contamination risks, potentially compromising data integrity. (4) Stable experimental conditions are needed for reliable results, limiting flexibility in varying environments. (5) Specialized systems may have limited throughput, constraining sample processing capacity. (6) The need for skilled technical expertise can create operational barriers, while (7) limited system flexibility and dependence on specialized equipment and intricate chip fabrication restrict innovation and customization. (8) Data privacy and security concerns on shared or cloud-based platforms may deter users, and (9) the system's complexity and high cost can limit access in resource-constrained environments.

In conclusion, these advancements in automation, cost-effectiveness, high-throughput capabilities, and enhanced precision have significant applications in molecular diagnostics, genomics research, and personalized medicine, enabling accurate disease monitoring and epidemiological tracking. However, addressing limitations such as connectivity and software compatibility issues[146,147,152], limited throughput[143,151], cross-contamination risks[40,143,149], the need for skilled expertise[29,143,146,152], and data privacy concerns is crucial for maximizing the potential of hardware and software integration in dNAAT systems. Ongoing research and development efforts should focus on overcoming these challenges to enhance the system's utility in molecular diagnostics, genomics research, and personalized medicine.

## 7. Outlook

This review has navigated the advancements in dNAAT, the intricacies of AI-enabled dNAAT image analysis, and their transformative potential. **Fig. 6** encapsulates the dynamic evolution of dNAAT technologies, with a particular emphasis on the transformative role of AI in advancing the fields of precision medicine and molecular diagnostics. This figure is divided into three pivotal phases, each marking significant milestones and future directions in the development and application of these technologies. In Phase 1, Invention and Early Development traces the origins of PCR technologies, from the groundbreaking invention of the first-generation PCR by Kary Mullis in 1983, through the advent of the second-generation qPCR in the mid-1990s, to the conceptualization of dPCR by Dr. Bert Vogelstein in 1999. This phase sets the foundation for the subsequent innovations in NAAT. In Phase 2, Commercial Development and Advancements focuses on the period from the early 2000s to the present, highlighting the introduction and commercialization of significant technologies such as LAMP in 2000 and its digital counterpart, digital LAMP, in 2011. This phase also covers the release of key platforms like the Fluidigm BioMark™ HD System and the QX100™ Droplet Digital PCR system, alongside the integration of AI to enhance the performance and capabilities of these technologies. In Phase 3, Integration and Widespread Adoption looks forward to the near future (2023-2024), anticipating the seamless integration of AI optimization with clinical applications, leading to the widespread adoption of these advanced systems. This phase underscores the potential of AI-enhanced dNAAT systems in revolutionizing point-of-care diagnostics and personalized medicine, facilitated by significant improvements in hardware, software, and the development of comprehensive sample-to-answer workflows.

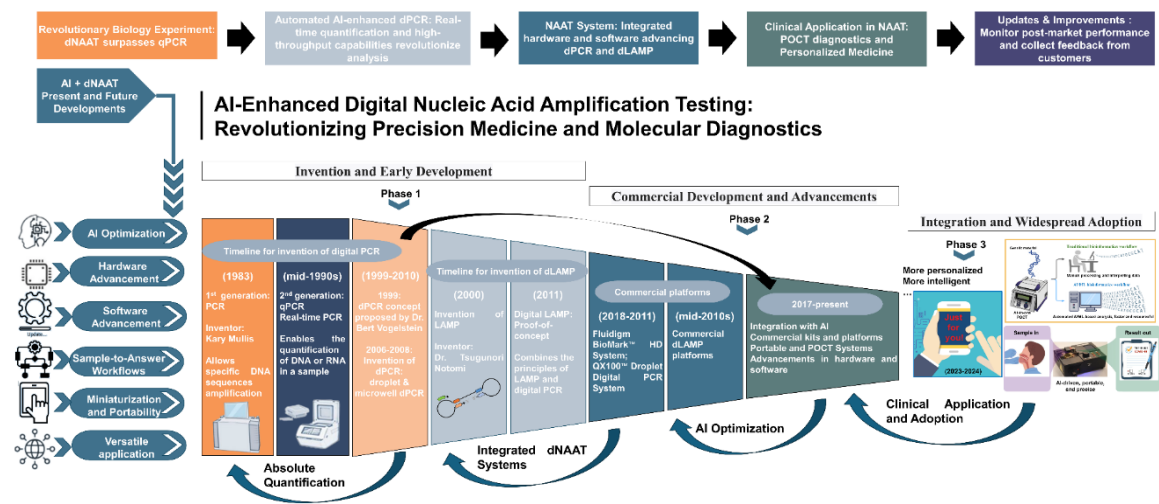

**Fig. 6 Conclusion and future outlook of AI-enhanced dNAAT technologies.** Key advancements and projected future directions for dNAAT technologies are depicted, underscored by the integration of AI. The timeline and schematic panels trace significant milestones, beginning with the inception of first-generation PCR platforms in the 1980s. The progress went through the advancements in second- and third-generation PCR and LAMP methodologies, culminating in the contemporary era of AI-augmented nucleic acid detection technologies poised to revolutionize biomedical diagnostics. Continuous advancements in AI and hardware technology are also accentuated, promoting optimized, high-throughput, and multiplexed analytical capabilities. The figure underscores the interplay between technological innovation and clinical application, highlighting the shift towards more personalized, intelligent diagnostic systems. It illustrates the critical role of AI in driving the evolution of dNAAT technologies, promising a future where diagnostics are not only more precise but also more accessible and tailored to individual patient needs. Ultimately, these innovations hold promise for profoundly impacting global health and biosecurity.

The convergence of AI with digital nucleic acid detection technologies marks a vanguard in life sciences. AI-driven enhancements in dNAAT protocols improve efficiency, accuracy, and cost-effectiveness, facilitating the development of automated, high-throughput systems that minimize human error. AI not only augments existing methods but also transforms data interpretation, leading to systems with superior efficiency, reliability, and capabilities. These AI-driven insights dismantle traditional barriers of time and cost, broadening the application scope of digital nucleic acid detection across genomic and diagnostic arenas[158,159]. Integrated dNAAT systems, leveraging SOTA hardware and software technologies for enhanced sensitivity and accessibility, herald a transformative future for preemptive diagnostics, particularly in resource-constrained regions. AI's adaptability to varied experimental conditions and its capability to process extensive datasets help achieve remarkable efficiency in both diverse and high-throughput applications.

While the prospects of AI-enhanced dNAAT are promising, several challenges need addressing to fully harness its capabilities and ensure broad adoption[160,161]. On the dNAAT side, cost and infrastructure pose significant hurdles; despite dNAAT's precise detection, the expense and infrastructure for these advanced systems can be daunting[162,163]. Efforts are underway to develop more affordable microfluidic devices and scalable AI algorithms that require less computational power, thus lowering costs[164,165]. Moreover, dNAAT is susceptible to errors, particularly in droplet interpretation, which can lead to inaccuracies[166,167]. Implementing internal positive controls is vital, as is training advanced AI models on diverse datasets to improve diagnosis reliability[64,71,163]. Challenges from the AI perspective include:

1. Cost and infrastructure: Initial investments for AI-native dNAAT systems are substantial, requiring robust computational infrastructure and specialized hardware[160].

2. Integration complexity: Merging AI with dNAAT demands meticulous system design and validation to ensure accuracy and reproducibility[168,169]. Simplifying integration through user-friendly platforms and automated processes is key to reducing complexity[10,39,169].

3. Data management and ethics: Handling the vast amounts of sensitive data generated by AI-driven dNAAT systems raises significant privacy and security concerns[170,171]. Encrypted databases and secure cloud services, coupled with strict regulatory compliance, are essential[134,169,172].

4. Algorithmic bias and error: Minimizing biases and errors in AI algorithms is crucial, especially for diagnostic systems that influence clinical decisions[31,173].

5. Regulatory compliance and standardization: Standardizing procedures for AI-enhanced diagnostics is challenging due to the rapid technological evolution and varied regulatory landscapes[173,174].

To overcome these challenges and unlock the full potential of AI-enhanced dNAAT, innovative solutions and strategic advancements are imperative.

1. Optimization and standardization of dNAAT protocols: Continual refinement of dNAAT assays and protocols through AI can vastly improve the accuracy and efficiency of nucleic acid amplification and detection[15]. AI algorithms can help optimize reaction conditions in real time, thereby ensuring consistent performance[134,175].

2. Large Language Models (LLMs) in analysis and synthesis: LLMs can analyze vast amounts of literature to identify prevailing issues and solutions in nucleic acid amplification, ensuring that researchers are informed of the latest advances and challenges[176,177]. Utilizing LLMs for image analysis algorithms can improve droplet identification accuracy, thus enhancing the overall sensitivity and specificity of dNAAT systems.

3. Robust training data sets: Design comprehensive training data sets that encompass a broad range of test conditions to help AI systems better generalize and perform with high accuracy under diverse real-world conditions. Ensure databases used for training AI models are inclusive of diverse genetic backgrounds to minimize biases and improve the models' utility across different populations. AI systems must be trained on appropriately matched datasets and continuously validated against empirical benchmarks to prevent overfitting and confirm their reliability[178]. Quantifying uncertainties in AI-powered analyses is also crucial.

4. Cross-disciplinary collaboration: Foster collaborations across bioinformatics, computer science, molecular biology, and clinical fields to develop standardized and efficient dNAAT protocols guided by AI[179,180].

5. Transparent AI models: Develop models with interpretable decision-making processes to build trust among users and facilitate regulatory approval[181,182].

6. Robust validation frameworks: Establish rigorous validation frameworks for AI-native dNAAT systems that parallel clinical testing benchmarks, ensuring they meet healthcare standards[183,184].

7. AI-Driven customization for local health needs: Leverage AI to tailor dNAAT systems to specific local health needs, particularly in resource-constrained environments, to enhance accessibility and impact[185]. Moreover, highly customized dNAAT solutions can be developed to match the specific needs of various applications such as rare disease detection or pathogen monitoring[186].

Looking ahead, we anticipate a dual trajectory: one path tackling dNAAT challenges beyond current technological scope through AI, and another steering towards AI-native designs of nucleic acid detection systems, built from the ground up with AI insights. These designs integrate multiplexing options, facilitating parallel assays and leveraging reduced sample volumes, setting the stage for portable diagnostics in resource-limited settings. The potential of AI-enhanced dNAAT is undeniable, and as we move forward, the incorporation of these advancements will lead to a more robust, accessible, and efficient molecular diagnostic framework. Through the amalgamation of cutting-edge AI with dNAAT, laboratories and medical professionals can expect a future where precision medicine isn't just a possibility—it's the norm.

Ultimately, the synergy of AI and digital nucleic acid detection technologies stands as a beacon for life sciences. AI elevates the capabilities of dNAAT and related methodologies, transforming the interpretation of complex datasets. These insights promise to reduce the time and financial barriers inherent in life science research, vastly extending the practical scope of digital nucleic acid detection technologies across genetic analysis and biomedical diagnostics.

## 9. Acknowledgements

The authors are grateful for the funding support from the Hong Kong Research Grants Council (project reference: GRF14204621, GRF14207920, GRF14207419) and the National Natural Science Foundation of China (Grants No.62204140). The authors would like to thank BioRender.com for providing the tools used to create the illustrations in this article.

## 10. Author information

Authors and Affiliations

**Department of Biomedical Engineering, The Chinese University of Hong Kong, Shatin, Hong Kong SAR, 999077, China**

Yuanyuan Wei, Wu Yuan and Ho-Pui Ho

**Department of Neurology, David Geffen School of Medicine, University of California, Los Angeles, California, 90095, USA**

Yuanyuan Wei

**Guangdong Institute of Intelligence Science and Technology, Hengqin, Zhuhai, 519031, China**

Xianxian Liu, Mingkun Xu

**Department of Computer and Information Science, University of Macau, Macau SAR, 999078, China**

Xianxian Liu

**Department of Computer Science Engineering, The Chinese University of Hong Kong, Shatin, Hong Kong SAR, 999077, China**

Changran Xu

**Department of Automation, Tsinghua University, Beijing, 100084, China**

Guoxun Zhang

**Center for Brain-Inspired Computing Research (CBICR), Department of Precision Instrument, Tsinghua University, Beijing, China**

Mingkun Xu

Contributions

Y. Wei and M. Xu conceived and designed the review theme; Y. Wei, X. Liu, and M. Xu gathered and analyzed the literature; Y. Wei, X. Liu, and M. Xu drafted the manuscript; Y. Wei and X. Liu created the figures; X. Liu created and Y. Wei revised the tables; M. Xu, C. Xu, and G. Zhang critically revised the manuscript for important intellectual content; M. Xu and H. Ho oversaw the project and acquired funding. All authors discussed the results and implications and commented on the manuscript at all stages.

Corresponding author

Correspondence to Ho-Pui Ho and Mingkun Xu.

## 11. Ethics declarations

No conflict of interest

## 12. Additional information

**Table 1. Comparative analysis of commercial dPCR platforms.**

| Manufacturer | Bio-Rad | Fluidigm | Thermo Fisher | RainDance | JN MEDSYS | Stilla Technologies | OPTOLANE |
|---|---|---|---|---|---|---|---|
| Function | 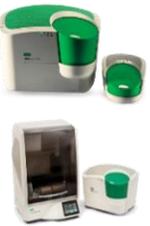 Droplet generator and droplet reader. | 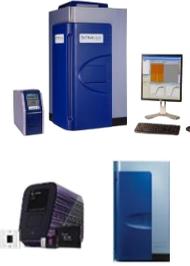 Biomark HD genetic analysis system. | 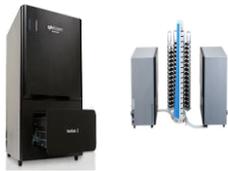 QuantStudio 3D (QS3D) genomics and molecular biology system. | 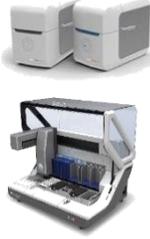 RainDrop™ genomics and molecular biology system. | 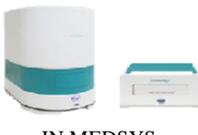 JN MEDSYS Clarity™ digital PCR system. | 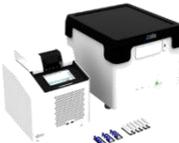 Crystal genomics and molecular biology system. | 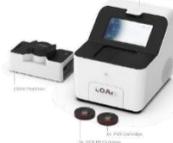 LOAA droplet dPCR system. |
| Platform (key product) | QX200 ddPCR System cat# 1864001; QX200 AutoDG ddPCR System cat #1864100. | Biomark™ HD System cat# cat. no. PN 100-5744; Juno™ & Biomark™ HD System cat# PN 100-7070. | QuantStudio Absolute Q dPCR System cat# A52864; QuantStudio Absolute Q AutoRun dPCR Suite cat# A57609-A57612. | RainDrop Plus™ dPCR System cat #20-04400; RainDance ThunderStorm™ System cat# no. 200315. | Clarity™ dPCR System cat # 10001. | Naica® system for Crystal dPCR™. cat #NAICA-06. | LOAA Analyzer cat# (RUO: LAM301-01-R) |
| Software | QX Manager Software; QuantaSoft™ Software and other compatible analysis software (QX200). | Microfluidics Software: Including Real-Time PCR Analysis and Genotyping Analysis software for data analysis from the Biomark HD system. | QuantStudio Absolute Q Software Version 6.3. | RainDrop® Data Analysis Software. | Clarity Chromatography Software; Clarity Elemental Analysis Software. | Crystal Miner. | LOAA Analyzer, (model name: On-Point); LOAA Sample Loader (model name: POSTMAN); |
| **Performance metrics** ||||||||
| Detection method[a] | End-point | Real-time | End-point | End-point | End-point | End-point | Real-time |
| Detection mode | Droplet plate | Dynamic Array™/ Digital Array™ integrated fluidic circuit (IFC)[100] | Microfluidic array plate (MAP) | Droplet chip | Chip-in-a-tube | Microfluidic chips | Genotizer™ 20K Well Chip, and D_PCR Chip Case[101] |
| Partitions | 20,000[102] | 36,720 | 20,000 | 5,000,000 to 10,000,000 | 10,000 to 40,000 (Clarity Plus™ Digital PCR system)[103] | 10,000 to 30,000 | 20,163 |
| Throughput | >96 samples per run. | 144 to 9,216 reactions per run. | 12 to 48 samples per run[104]. | 1 to 96 samples per run[100]. | 96 samples per run. | 12 to 48 samples per run[105]. | - |

[a] **Definition:** Endpoint PCR is a method where DNA amplification proceeds to completion, with analysis occurring post-process. It detects and quantifies DNA following all PCR cycles, justifying the "endpoint" label. In contrast, Real-time PCR (qPCR) observes DNA amplification in real-time using fluorescent markers, allowing DNA quantification at each cycle through fluorescence measurement.

| Manufacturer | Bio-Rad | Fluidigm | Thermo Fisher | RainDance | JN MEDSYS | Stilla Technologies | OPTOLANE |
|---|---|---|---|---|---|---|---|
| **Duration (the total duration from sample preparation to final results)** | ∼4 to 5h[106]. Sample preparation: ~30 mins to 1h; Droplet Generation: 5 to 10 minutes per 8 samples in one droplet generator cartridge; PCR amplification: 2h; Droplet reading: 1h; Data analysis: <1h. | ∼4 to 6h. Sample preparation: ~30 mins to 1h; Chip priming and loading (IFCs loading) ~ 30 mins to 1h; Thermal cycling: 1.5 to 2h; Data collection and analysis: ~1h. | ∼3.5 to 4h. Sample preparation: 1 to 2h; Plate setup: 15 to 30 mins; Thermal cycling: ~2h; Reading and data analysis: 30 mins to 1h. | ∼3 to 4.5h. Sample preparation: 1 to 2h; Droplet generation: 5 to 10 mins per sample; Thermal cycling: ~ 2h; Droplet reading: 15 to 30 mins per sample; Data analysis: 15 to 30 mins. | ∼3.5 to 4.5h. Sample preparation: ~1 to 2h; Reaction setup: 15 to 30 mins; Thermal cycling: 2h; Reading and data analysis: 30 mins to 1h. | ∼3.5 to 4.5h. Sample preparation: 1 to 2h; Reaction setup and chip loading: 15 to 30 mins; Thermal cycling: 2h; Imaging and data analysis: 30 mins to 1h. | ∼3.5 to 4.5h. Sample preparation: 1 to 2h; Reaction setup: 15 to 30 mins; Thermal cycling: 2h; Reading and analysis: 30 mins to 1h. |
| **Reaction volume (µL)** | 20 | 117 to 252 for 60 reactions | 15 | 25 - 50[107] | 15[108] | - | 30 |
| **Number of channels** | 2-plex | 2-plex | 2-plex | 2 to 5-plex | 6 to 8-plex | - | 2-plex |
| **Detection channels** | 2-colors: FAM (EvaGreen), HEX (VIC) (QX200)[102]. 6 colors: FAM, HEX, Cy5, Cy5.5, ROX, and ATTO 590 (QX600)[109]. | 2-colors. | - | 2-colors FAM (512 nm) and VIC (543 nm). | 2-colors: FAM™, VIC/HEX™ (Clarity™ dPCR System). 6-colors: FAM™, HEX™, ATTO™ 550, Texas Red®, Cy5™, and Cy5.5™ (Clarity Plus™ dPCR system). | 3-colors. | 3-colors. |
| **Time per run** | ∼2h to 4h. | ∼2h to 4h. | ∼1h to 4h. | ∼2h to 4h. | - | ∼2h to 4h. | - |
| **Partition vol (nL) each droplet** | 0.85[110]. | 0.48 for dPCR 37k chips to 6 for 12.765 chips. | 0.865. | ∼0.005. | ∼1.336 for Clarity™. 0.3 for Clarity Plus™. | - | - |
| **Cost info** | | | | | | | |
| **Platform cost (euros)** | 70,000[111] | ~600,000 to 700,000 | ~30,000 to 40,000 | ~110,000 | - | ~120,000 | ~17,573 |
| **Application** | | | | | | | |
| **Key applications** | Rare mutation detection (ctDNA, (CTCs), and other biomarkers present in blood, urine, or other bodily fluids. | SNP genotyping; High-throughput SNP analysis. | Rare mutations in circulating tumor DNA (ctDNA). | GWAS follow-up; Deep sequencing of heterogeneous tumor samples; SNP; CNV; DSG. | Detection and quantification of viral and bacterial pathogens; Detecting rare mutations and quantifying oncogene expressions; Liquid biopsy. | Liquid biopsy and non-invasive prenatal testing (NIPT) for circulating tumor DNA (ctDNA) and fetal DNA in maternal blood. | Liquid biopsies (cfDNA, CTC); NGS library quantification /validation. |

**Table 2. Comparative evaluation of early-stage dPCR platforms.**

| Manufacturer | QIAGEN, Hilden, Germany | Life Technologies (integrated into Thermo Fisher Scientific) | SlipChip[112] | Sysmex Inostics | Stilla Technologies | Rubicon Genomics | Leiden University Medical Center |
|---|---|---|---|---|---|---|---|
| **Platform and function** | 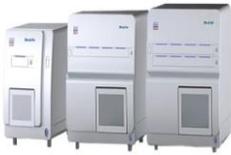 | 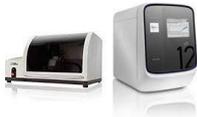 | 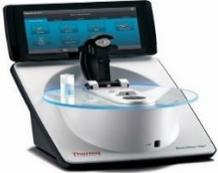 | 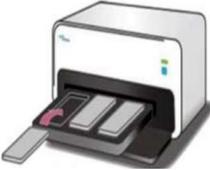 | 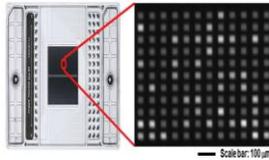 | 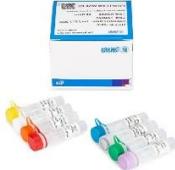 | NA |
| | QIAcuity™ genomics & molecular biology system[113,114]. | OpenArray Digital PCR System. For Use with QuantStudio™ 12k Flex | | BEAMing Digital PCR System[115]. | METEOR dPCR System[116]. | PicoPLEX WGA digital PCR. | Genomic Microfluidic Digital PCR (GMDP) digital PCR |
| **Performance Metrics** | | | | | | | |
| **Detection mode** | End-point & Real-time | Real-time[117,118] | End-point[119] | End-point | End-point[120] | End-point | Real-time[121] |
| | Nanoplate well[114] | Microfluidic chip | Microfluidic chip[122] | Droplet chip[123] | Droplet chip | Microfluidic chip[124] | Microfluidic chip. |
| **Partitions** | 8,500[114] | - | 20,000[112,121] | - | - | 10,000 - 30,000 | 10,000 - 30,000 |
| **Detection channels** | Two or more colors[125]. | 6 colors (21 filter combinations). | - | - | - | 3-color. | 3-color. |
| **Time per run** | ~2 to 6 h. | ~2 to 4h. | ~1 to 4h. | Several hours to a day or more. | - | ~2 to 4h. | - |
| **Automation** | Available. | Available. | NA | Some integrated systems are available for automated sample preparation, emulsion PCR, emulsion breaking, and analysis. | The integrated system is partially automated in liquid handling, emulsion generation, emulsion breaking, detection, and analysis. | NA | - |
| **Application** | | | | | | | |
| **Key application and use cases** | Rare mutations detection; Low-abundance targets quantification; Nucleic acids quantification. | Nucleic acid quantification; Protein quantification; Enzyme assays; Cell culture applications; Environmental and food analysis. | Screening and early detection; Real-time monitoring of therapy; Evaluation of early treatment response; Monitoring of minimal residual disease; Risk for metastatic relapse | Mutation detection; Profiling; Infectious disease diagnostics; Cancer research and liquid biopsy; Genetic testing and personalized medicine. | Single-cell genomics; Clinical research and diagnostics; Cancer research; Microbiome studies; Forensic science; Ancient DNA analysis; Basic research. | Mutation detection. Liquid biopsy. | Infectious disease diagnostics; Forensic genetics; Prenatal testing; Pharmacogenomics; Environmental and agricultural genomics. |

| Manufacturer | QIAGEN, Hilden, Germany | Life Technologies (integrated into Thermo Fisher Scientific) | SlipChip[112] | Sysmex Inostics | Stilla Technologies | Rubicon Genomics | Leiden University Medical Center |
|---|---|---|---|---|---|---|---|
| | | | (prognostic); Patient stratification; Mechanisms of therapeutic targets and resistance. | | | | |

**Table 3. Overview of AI-enhanced dNAAT advances.**

| Application | Learning type | Algorithm architecture | Performance Metrics | Dataset information | Bio experiment set-up | Assessment[b] |
|---|---|---|---|---|---|---|
| Digital microwell PCR-HRM for bacterial identification[140] 2023 | Machine learning /Supervised | Supervised: ovo-SVM | Accuracy: ~100%. Generality: Emphasizes capability for multiple bacterial species identification. Accessibility: UmeltSM via https://dna-utah.org/umelt/quartz/. | Employed digital melt curves from five bacterial species (Train: test =8:2). | Utilized custom nanoarray chips with a CFX Real-Time PCR system for bulk-based melt curve generation. | Strength: 1, 2, 3, 4) *, (6) #. Limitation: (4) *. |
| Automated droplet classification for BRAF V600E[136] 2021 | | Supervised: k-NN | Sensitivity: Comparable sensitivity to manual selection. Accessibility: Leveraged R software for accessible GUI development. | Examined from 73 microPTC cases and 115 thyroid specimens. | Conducted using the Bio-Rad QX100 System, combining a TaqMan assay for BRAF V600E quantification. | Strength: (1, 3, 4) *; (6) #. Limitation: (1, 2) *. |
| eDNAssay: Predictive qPCR cross-amplification[141] 2022 | | Random forest classifiers | Accuracy: Primer-only model at 92.4% accuracy; Full-assay model reaches 96.5%. Generality: Tolerance of some polymorphisms within target taxa and target polymorphisms. Accessibility: eDNAssay via https://NationalGenomicsCenter.shinyapps.io/eDNAssay. | 530 assay–template combinations: 262 with primers alone and SYBR Green intercalating dye, 268 with primers and TaqMan hydrolysis probes. | Validated using StepOnePlus and QuantStudio 3 Real-Time PCR Systems. | Strength: (2, 3) *; (6) #. Limitation: (1) *; (7) #. |

[b] **Assessment rubric:**

**Strengths:** (1) Highly automated; (2) Versatile for diverse research and experimental setups and even various applications; (3) Enhances sensitivity and precision; (4) Time-saving, user-friendly, reduces repeat testing and downstream confirmatory tests, enhancing cost-effectiveness and real-time processing; (5) Efficient learning from limited data without extensive data collection and training; (6) Supporting personalized medicine; (7) Multiplexing for molecular diagnostics and quantitative analysis; (8) Supports public health decisions.

**Weaknesses:** (1) Model interpretability and transparency issues; (2) Dependence on high-quality data and risk of overfitting; (3) Scalability concerns for larger datasets or complex analysis tasks with limited data; (4) Generalization challenges that limited to specific molecular techniques or applications; (5) Limited accessibility and usability for non-expert users; (6) Potential cross-reactivity in multiplex assays; (7) Data privacy and security concerns on shared/cloud platforms; (8) Limited to color-based differentiation and alternative multiplexing strategies.

Strengths 1-4 (*) highlight universally recognized benefits, while Strengths 5-8 (#) denote the unique advantages of custom designs. Conversely, Weaknesses 1-5 (*) signify common challenges across algorithms, suggesting areas for widespread enhancement. Weaknesses 6-8 (#), specific to the methodologies, pinpoint precise refinement needs for improved system performance and reliability.

| Title | Category | Method | Metrics | Dataset | Platform | Strength/Limitation |
|---|---|---|---|---|---|---|
| Deep-qGFP: droplet and microwell dPCR detection and quantification[39] 2023 | Deep learning /Supervised | Supervised: Yolov5 with RPN | Accuracy: 96.23%. Throughput: 2,000 droplets in 2.5 seconds. Sensitivity: 56.52 to 1569.43 copies µL$^{-1}$. Calibration: $R^2$ = 0.9433 for droplet; $R^2$ = 0.7860 for microwell. Generality: Demonstrated droplet dPCR, microwell dPCR, agarose dPCR, and other droplet-based applications. Accessibility: Leveraging Python for GUI development. | 206 concentration-varying images for labeling. | Implemented a customized microfluidic system using a flow-focusing chip. | Strength: (1, 2, 3) *. Limitation: (1, 2, 3) *. |
| Accurate detection and classification of droplets from low-quality ddPCR images[139] 2023 | | Supervised: Attention DeepLabV3+ combined with CHT | Accuracy: >97.0 %. Throughput: ~2 seconds per image of 512-by-512-by-3 pixels. Robustness: 25 epochs. | Comprised 24 dPCR images enhanced to 4224 through augmentation techniques, assigned weights for classification, and split into an 80-20 training-validation ratio for machine learning model development. | Amplified using BioRad system and quantified using Nanodrop1000 system. | Strength: (1, 2, 3) *. Limitation: (2, 3, 4) *. |
| SCAD: single fluorescent channel multiplex microwell dPCR detection[41] 2022 | | ViT combined with tSNE | Accuracy: 98.65%. Robustness: 150 epochs. Calibration: exceptional calibration with $R^2$ > 0.995 for both single-target and multiplex samples. Generality: widely applicable for various dPCR methods. | Analyzed 5000 images of single microwell; randomized training/validation split (7:3 ratio); blaNDM and blaVIM targets. | Deployed on a 3D Digital PCR chip v2; focused on multiplex detection challenges. | Strength: (2, 3, 4) *; (7) #. Limitation: (1, 2) *; (6, 7, 8) #. |
| Automatic localization and classification in ddPCR imaging[37] 2023 | | Supervised: Hough transform with CNNs | Accuracy: 95.04% for positive, and 99.71% for negative droplets. Throughput: Processing 2456 × 2404 pixels ddPCR image in 7 seconds. Robustness: Model training for 300 epochs. Sensitivity: 95.98% with 5-layer CNN and 99.62% with 3-layer CNN. | Valid droplets: 224; Invalid droplets: 3168; Positive droplets: 4124; Negative droplets: 8570. | Sapphire microfluidic chip; Stilla NaicaPrime3 imaging. | Strength: (2, 3, 4) *; (7) #. Limitation: (1, 2) *. |
| Accurate detection and quantification in droplet dPCR[137] 2023 | | Supervised: Mask R-CNN with GMM | Accuracy: > 93% across homogenous and non-homogenous. Robustness: Model training within 2,000 iterations (epoch). Calibration: $R^2$ = 0.9973. Generality: Applicable for various infectious diseases. | Utilized 20 high-resolution images, augmented to 64 for training and 16 for validation. | Focused on E. coli O157:H7 DNA; utilized one-step RocketScript™ Reverse Transcriptase and quantification via a droplet-based chip. | Strength: (1, 3, 4) *; (7) #. Limitation: (2, 5) *; (7) #. |
| YOLOv5 & CBAM for droplet identification[138] 2023 | | YOLOv5 with CBAM | Accuracy: 98%. TP:2735, FP:39, TN:3531, FN:1. Throughput: Completes analysis of 2735 droplets in 2 seconds. Robustness:5000 epochs in model training. Sensitivity: TP:2735, FP:39, TN:3531, FN:1. Accessibility: accessible via mobile devices or cloud platforms. | 120 images are obtained from 40 images by reducing the brightness of these images or applying Gaussian noise. randomized training/validation/testing split (7:2:1 ratio). | Utilized the QX200 Droplet Generator and a custom microscopic imaging platform for high-resolution droplet analysis. | Strength: (1, 2, 3, 4) *. Limitation: (1, 2, 5) *. |
| Novel few-shot droplet dPCR detection by YOLOv3, RBTM&STAM[38] | Clustering /Unsupervised | Few-shot learning adpted YOLOv3 | Accuracy: 98.98%. Throughput: Reduces labeling time by over 70%; processing speed of 5.25 times/second, 4.5 times faster than Mask R-CNN and YOLOv3 individually. | Analyzed 120 training and 40 testing images from chip-based and 80 training and 20 testing micro-droplet dPCR setups, explored through data | Acquired images from reported works, utilizing an integrated ddPCR chip with the configuration of flow-focusing droplet | Strength: (1, 2, 3, 4) *; (5) #. Limitation: (2, 3, 4) *. |
| 2021 | | | | augmentation techniques for in-depth image analysis. | generator and droplet array chamber. | |

**Table 4. Overview of integrated dNAAT system advances.**

| System name | Function | Innovation | Setup (Hardware & Software) | Performance Metrics | Bio experiment Setup & Dataset information | Evaluation[c] |
|---|---|---|---|---|---|---|
| Smartphone-integrated handheld (SPEED) digital polymerase chain reaction device[99]. 2023 | High-throughput 8-plex ddPCR system for DNA methylation detection. | Smartphone-integrated handheld dPCR. | Hardware: Smartphone (Huawei P40)-integrated handheld dPCR. Compact (100 mm x 200 mm x 35 mm) and lightweight (400g). Integrated, miniaturized modules for thermal cycling, image taking, and wireless data communication. Consumable is the developed silicon-based dPCR chip. Software: Controlled by self-developed Android-based applications. MATLAB mobile app analytics. Software controls temperature cycling, as well as LED illumination of the chip and cooling fan. | Precision and accuracy are comparable with commercial dPCR machines. Throughput: 45 PCR cycles in ~49 minutes. silicon-based dPCR chip dimensions of $9 \times 9$ mm$^2$ and 26,448 partitions. The camera has 12.6 M pixels, providing ~50 pixels per well. | Verified in the testing of severe acute respiratory syndrome coronavirus 2, the detection of cancer-associated gene sequences, and the confirmations of Down syndrome diagnoses. | Strength: (1, 4) *; (7) #. Limitation: (1) *. |
| Deep-dLAMP for label-free, low-cost nucleic acid quantification[36]. 2022 | Label-free, cost-effective nucleic acid quantification. | Polydisperse emulsion-based digitization incorporating deep learning. | Hardware: Off-the-shelf lab hardware including vortex mixer, thermal cycler, and camera-coupled brightfield microscope. Software: Mask R-CNN (image segmentation) with Resnet-50; SGD optimization & imgaug augmentation for deep learning image analysis (emulsion segmentation, volume regression, and precipitate-based occupancy classification). | Accuracy: 88% mAP with IoU up to 0.8. Sensitivity: Limit of detection of 5.6 copies µl$^{-1}$. Dynamic range: 37.2 to 11000 copies µl$^{-1}$. | Proteus mirabilis and SARS-CoV-2 cases. The dataset comprises a total of 638 image frames at different nucleic acid concentrations. | Strength: (1,5) *; (6,11) #. Limitation: (1,2) *; (3) #. |
| StratoLAMP for label-free and multiplex dLAMP based on | Multiplex dLAMP system with visual precipitate | Eliminates the need for labeling reagents. Simultaneous quantification of two | Hardware: Droplet generation by a flow-focusing microfluidic chip with the flow driven by a negative pressure. Microfluidic chip fabricated using photolithography and replica | Accuracy: 94.3%. Calibration: R$^2$ = 0.99 in testing individual targets. | Validated using human papillomavirus 16 (HPV 16) and HPV 18 for three classes of | Strength: (2) *; (6, 8, 10) #. Limitation: (4, 6) #. |

---

[c] **Assessment rubric:**
**Strengths:** (1) Automated system integration. (2) Cost-effective and performance-enhancing hardware upgrades, including rapid thermal cycling et al. (3) Software advances for cost efficiency and large-scale screening. (4) Versatility in supporting various dPCR applications. (5) High processing capacity. (6) Multiple target analysis in a single run. (7) Portable solutions for POCT and underserved regions. (8) Elimination of labeling processes. (9) Low risk of cross-contamination. (10) Reduction in sample and reagent requirements. (11) Compatibility with complex samples. (12) Supports epidemiological studies. (13) Enhanced consistency in results.
**Limitations:** (1) Necessity for specialized hardware. (2) Compatibility issues with existing software and connectivity. (3) Increased contamination risk with miniaturized and capillary systems. (4) Dependence on stable conditions for reliable outcomes. (5) Throughput constraints in specialized systems. (6) Need for specific technical skills. (7) Rigid reliance on specialized devices and intricate manufacturing. (8) Data security concerns in shared environments. (9) Accessibility challenges due to system complexity and cost.
Strengths 1-5 (*) highlight universally recognized benefits, while Strengths 6-13 (#) denote unique advantages from custom designs. Conversely, Weaknesses 1-2 (*) signify common challenges across algorithms, suggesting areas for widespread enhancement. Weaknesses 3-9 (#), specific to the methodologies, pinpoint precise refinement needs for improved system performance and reliability.

| Device | Purpose | Innovation | Hardware/Software | Performance | Application | Strengths/Limitations |
|---|---|---|---|---|---|---|
| visual stratification of precipitate[40]. 2024 | stratification for label-free detection. | nucleic acid targets using the inherently generated precipitate in isothermal amplification as the detecting feature. | molding. Thermal cycler for amplification. An inverted fluorescence microscope coupled with a camera for imaging. Software: A Mask R-CNN model using ResNet-50 as the backbone and region proposal networks to scan for droplets within the images. | Dynamic range: $2.46 \times 10^1$ to $6.90 \times 10^3$ copies $\mu l^{-1}$. Accessibility: Point-of-care and field-testing adaptable design. | precipitate quantification by setting each set of primers to proper concentrations. | |
| An integrated ddPCR lab-on-a-disc (LOAD) device for rapid screening of infectious diseases[51]. 2023 | Rapid screening of infectious diseases in various settings, including point-of-care and field applications. | The LOAD device integrates sample preparation, nucleic acid amplification, and detection in a single platform. It uses centrifugal microfluidics to automate the process. | Hardware: Integrated microfluidics chips, transparent circulating oil-based heat exchanger, and centrifugal force for droplet generation. Software: Automated control of sample preparation, nucleic acid amplification, and detection. | Sensitivity: Limit of detection of 20.24 copies $\mu l^{-1}$. Accessibility: Point-of-care and field-testing adaptable design. | Tested for five different viruses, including influenza A virus (IAV), respiratory syncytial virus (RSV), varicella zoster virus (VZV), Zika virus (ZIKV), and adenovirus (ADV). | Strength: (1, 4) *; (7) #. Limitation: (1) *; (6) #. |
| Capillary-based integrated ddPCR[29]. 2018 | Direct quantification with zero droplet loss/no cross-contamination. | First capillary-based ddPCR integrating droplet formation, cycling, and detection. All-in-one scheme. | Hardware: An HPLC T-junction is used to generate droplets and a long HPLC capillary connects the generator with both a capillary-based thermocycler and a capillary-based cytometer. Software: A single-chip microcontroller (ARDUINO UNO REV3) coded with PID control assisted by thermocouples for temperature control. | Calibration: linearity at $R^2 = 0.9988$. Dynamic range: NTC to $2.4 \times 10^{-4}$ copies $\mu l^{-1}$. | Lung cancer gene LunX as the target. | Strength: (1, 5) *; (7) #. Limitation: (6, 7, 8, 9) #. |
| Self-priming compartmentalization (SPC) integrated dPCR[145]. 2014 | 'Divide and conquer' single molecule detection via on-chip valve-free microfluidics. | Simplifies compartmentalization through self-priming control. The first integrated on-chip valve-free and power-free microfluidic digital PCR device. | Hardware: Rectangular microwells with main channels for sample division. The SPC microfluidic chip (50 mm x 24 mm x 4 mm) contains 5120 independent 5 nL microchambers, fabricated with multilayer soft lithography techniques. Software: The chip after amplification was detected using the Maestro Ex IN-VIVO Imaging System (CRI Maestro). Fluorescence images were acquired by using a large-area CCD system. | Accuracy: < 5% measurement deviation. Calibration: $R^2 = 0.995$, $R^2_{PLAU} = 0.998$, $R^2_{ENO2} = 0.999$, $R^2_{PLAT} = 0.999$. Lifespan: SPC chip sealed with transparent adhesive tape could last for ~4 h. | Three lung cancer-related gene abundance analyses. Usage of Maestro Ex IN-VIVO Imaging System (CRI Maestro). | Strength: (1, 5) *; (9, 11) #. Limitation: (6, 10) #. |
| Raspberry Pi-based integrated real-time dPCR system[25]. 2019 | Elevated sensitivity and validation in microwell real-time dPCR analysis. | Raspberry Pi-based cost-efficient integration for in-array real-time detection. | Hardware: A highly integrated real-time dPCR device combining a thermocycler, a fluorescence image setup, a Raspberry Pi with a touch screen, and a microfluidic dPCR chip. Software: A Linux-based application for Raspberry Pi was developed in C++ based on Qt Creator for real-time fluorescence image acquisition, image processing, and image data analysis. | Calibration: $R^2 = 0.9998$. Dynamic range: $1 \times 10$ to $2 \times 10^3$ copies $\mu l^{-1}$. | Permitted higher quantitative accuracy than the commercial Quantstudio™ 3D dPCR system when assaying a plasmid DNA containing the human 18S ribosomal RNA gene fragment. | Strength: (1, 4) *; (9) #. Limitation: (1, 2) *. |

| Device | Description | Key feature | Hardware/Software | Performance | Application | Strength/Limitation |
|---|---|---|---|---|---|---|
| Fully automatic integrated continuous-flow dPCR[144]. 2020 | Fully automated dPCR system based on continuous-flow theory. One-step operation with complete control and data processing software integration. | Fully automatic integrated dPCR based on continuous flow dPCR theory | Hardware: Sized 467 × 420 × 185 mm. Powered by solar cells and does not require an external power supply. Controlled by a notebook. Employs a self-designed fluorescence analysis system. Software: The host computer software is based on Python for complete control of one-step operation. The data processing software is based on openCV. | Calibration: $R^2 = 0.9986$. Dynamic range: three orders of magnitude from $10^3$ to $10^5$ IU/mL. | Validated through absolute quantitative detection of the hepatitis B virus in serum samples. | Strength: (1, 5) *. Limitation: (1, 2) *. |
| A handheld, smartphone-based mobile dPCR device[142]. 2018 | Cost-effective, portable dPCR device for highly accurate DNA analytics. | Mobile device integration with controls for thermal cycling, dPCR reaction, and data analysis. | Hardware: A handheld, smartphone-based mobile dPCR device integrated with thermal cycling control, on-chip dPCR. Compact (90 mm x 90 mm x 100 mm) and lightweight (500g). Under $320 excluding smartphone. Software: Customize Android software for automation. | Calibration: $R^2 = 0.9994$. Generality: Versatile genetic marker detection from diverse sources such as blood, saliva, or tissue samples. Accessibility: Cost-effective and customized Android software accessible. | Accurately quantifed down to 10 copies of the human 18 S ribosomal DNA fragment inserted in a plasmid. Detected single molecule of cancer biomarker gene CD147 in a low number of HepG2 cells. | Strength: (1, 4) *; (6, 7) #. Limitation: (4, 11) #. |
| Microfluidic array partitioning (MAP) consumable device for dPCR[142]. 2019 | Simplifies dPCR workflows and reduces contamination risks for swift clinical results. | Novel integrated platform with real-time and melt-based partition analysis. | Hardware: Micro-molded plastic microfluidic consumable with a fully integrated single dPCR instrument that combines pneumatic consumable sample loading, thermal cycling, and 3-color fluorescence image acquisition with control and analysis software. Software: Fully automated software controls the sample digitization into the MAP partitions, thermal cycling, and imaging. | Throughput: sample-to-answer in <90 minutes. Generality: Three unique cancer-specific applications. Efficacy: >95 % sample compartmentalization. | Cancer-specific applications for EGFR T790M rare mutant quantification and BCR-ABL1 rare transcript quantification. Personalized cancer monitoring using a patient-specific dPCR assay. | Strength: (1, 5) *; (9) #. Limitation: (6, 12) #. |
| Microfluidic impact printing (MIP) ddPCR[143]. 2020 | Microfluidic impact printing technology for sensitive detection of rare variants. | Non-contact droplet generation technique promoting accuracy and avoiding cross-contamination. | Hardware: Platform composes of a microfluidic printing device, a PCR amplification device and a fluorescence detection apparatus. Software: Image analysis using ImageJ. Image batch analyzed using MATLAB. | Calibration: $R^2 = 0.997$. Dynamic range: 0.464 to 464 copies $\mu l^{-1}$. Efficacy: >92.6 % sample compartmentalization. | Quantification of p53 gene in colon cancer tissues. | Strength: (1, 2, 4, 5) *; (10) #. Limitation: (6, 8, 9) #. |
| Real-time dPCR (RT-dPCR) integrated device[146]. 2022 | Supports real-time digital PCR analysis within an all-in-one device. | Low-cost integrated RT-dPCR device for advanced polymerase chain reaction analysis. An improved data classification method for RT-dPCR based on the amplification process of each subsample. | Hardware: A multi-well chip (~20,000 hexagonal microwells etched) made of silicon substrate was used for amplification. The interior and outer parts were processed with hydrophilic and hydrophobic treatments respectively, to improve the microwell filling rate. Software: Enhanced process-based classification model (PAM) for data classification. | Calibration: $R^2 = 0.97$. | Classification tasks for different concentrations of Epstein-Barr Virus (EBV) plasmid quantification assays. | Strength: (1, 2) *; (11) #. Limitation: (5, 9, 11) #. |

| Method | Key Feature | Design | Hardware/Software | Performance | Validation | Strength/Limitation |
|---|---|---|---|---|---|---|
| Multi-volume microfluidic device with no reagent loss for low-cost dPCR[147]. 2020 | Achieves complete sample compartmentalization without loss for precise detection. | Vertical branching microchannel chip for multi-level volume partitioning without manual intervention. | Hardware: The multi-volume microchannel dPCR chip with a glass-PDMS-glass "sandwich" structure is fabricated using the multi-layer soft lithography process. Software: Image analysis using ImageJ. | Calibration: $R^2 = 0.9997$. Dynamic range: $>10^4$. Efficacy: 100 % sample compartmentalization without any loss. | Validated by measuring a 10-fold serial dilution of the KRAS plasmid template. | Strength: (1, 2, 4) *; (10) #. Limitation: (5, 9, 10) #. |
| Pressurized microheater-based thermal cycler dPCR[152]. 2021 | Designed a silicon-based fast-generated static droplets array (SDA) chip and developed a rapid dPCR detection platform that is easy to load samples for fluorescence monitoring. | Integration of a silicon-based fast-generated open well SDA chip and a rapid dPCR detection platform with a pressurized elaborate water evaporation control system. | Hardware: The platform consists of a dPCR chip, a data acquisition system, and a pressurized thermal cycler system. The SDA chip is fabricated by photolithography and dry etching processing technology to fabricate 2704 independent microarrays (each volume of about 0.785 nL) on a 15 mm × 15 mm silicon substrate chip. The real-time fluorescence monitoring system included a blue-color LED light source, a set of filters, a set of lenses, and a CMOS camera. | Calibration: $R^2 = 0.9999$. Sample loading time: < 10 seconds. Dynamic range: 3.13 copies $\mu l^{-1}$ to $3.13 \times 10^4$ copies $\mu l^{-1}$. | Tested using a gradient dilution of the hepatitis B virus (HBV) plasmid as the target DNA for a dPCR reaction. | Strength: (1, 2, 4) *; (10) #. Limitation: (5, 7, 9, 12) #. |
| Vibrating sharp-tip capillary-based ultra-wide dynamic range dPCR[149]. 2021 | Realizes real-time modulation of droplet size for high throughput and monodispersity. | Unique droplet generation relies on acoustic streaming from a vibrating sharp-tip capillary, enabling efficient droplet generation without any external pressure sources. | Hardware: Droplet generation based on the acoustic streaming generated from a vibrating sharp-tip capillary, consisting of a pulled glass capillary, a microscope glass slide, and a piezoelectric transducer. The system can be driven by a battery-powered low-cost signal generator. Software: Image analysis using ImageJ. Poisson distribution calculation using dPCR MATLAB. | Throughput: Up to 5000 droplets/s. Calibration: $R^2 = 0.986$. Dynamic range: ~6 orders of magnitude. Generality: Multiplexed measurements of cancer biomarker HER2 gene and reference EIF5 gene in five different breast cancer cell lines. Monodispersity: >96%. Power consumption: <60 mW. | Validated by multiplexed measurements of cancer biomarker HER2 gene and reference EIF5 gene in five different breast cancer cell lines. It enabled a simple protocol for the non-invasive fabrication of cell-laden alginate microcapsules with variable sizes. | Strength: (1, 4) *; (7, 10) #. Limitation: (6, 7, 10) #. |
| A microfluidic alternating-pull–push active digitization (µAPPAD) for enhanced sensitivity[151]. 2019 | µAPPAD for reproducible and highly efficient digitization of samples of limited volume. | µAPPAD employs integrated pneumatic valves to periodically manipulate air pressure inside the channels to greatly facilitate the microwell-based fluid compartmentalization. | Hardware: Pneumatic valves within the microfluidic channels aid in digitization. Software: Image analysis using ImageJ. | Accuracy: 99.5 ± 0.3%. Efficacy: ~100% sample digitization efficiency. Dynamic range: ~7 to 3600 copies $\mu l^{-1}$. | λDNA as the model target using the tandem-channel chip. | Strength: 1 *; (10, 11, 13) #. Limitation: (5, 7, 8, 9, 12) #. |